\documentclass[aps,prl,floatfix,twocolumn,reprint,amsmath,amssymb,superscriptaddress]{revtex4-1}
\usepackage{amsfonts}
\usepackage{mathrsfs}
\usepackage{amsmath}
\usepackage{color}
\usepackage{natbib}
\usepackage{graphicx}
\usepackage{bm}
\usepackage{amssymb}
\usepackage{xspace}
\usepackage{epstopdf}
\usepackage{dcolumn}
\usepackage{multirow}
\usepackage[colorlinks=true, letterpaper=true, pdfstartview=FitV, linkcolor=blue, citecolor=blue, urlcolor=blue]{hyperref}

\makeatletter

\newcommand{\Rmnum}[1]{\expandafter\@slowromancap\romannumeral #1@}
\makeatother

\begin{document}
\title{Unconventional Pairing Induced Anomalous Transverse Shift in Andreev Reflection}

\author{Zhi-Ming Yu}
\affiliation{Research Laboratory for Quantum Materials, Singapore University of Technology and Design, Singapore 487372, Singapore}

\author{Ying Liu}
\affiliation{Research Laboratory for Quantum Materials, Singapore University of Technology and Design, Singapore 487372, Singapore}

\author{Yugui Yao}
\affiliation{Beijing Key Laboratory of Nanophotonics and Ultrafine Optoelectronic Systems, School of Physics, Beijing Institute of Technology, Beijing 100081, China}

\author{Shengyuan A. Yang}
\affiliation{Research Laboratory for Quantum Materials, Singapore University of Technology and Design, Singapore 487372, Singapore}

\begin{abstract}
Superconductors with unconventional pairings have been a fascinating subject of research, for which a central issue is to explore effects that can be used to characterize the pairing. The process of Andreev reflection---the reflection of an electron as a hole at a normal-mental-superconductor interface---offers a basic mechanism to probe the pairing.
Here we predict that in Andreev reflection from unconventional superconductors, the reflected hole acquires an anomalous spatial shift normal to the plane of incidence, arising from the unconventional pairing. The transverse shift is sensitive to the superconducting gap structure, exhibiting characteristic features for each pairing type, and can be detected as voltage signals. Our work not only unveils a fundamentally new effect with a novel underlying mechanism, but also suggests a possible new technique capable of probing the structure of unconventional pairings.
\end{abstract}
\maketitle

Interface scattering---the scattering at an interface between different media---is ubiquitous for all kinds of particles and waves. It offers a basic means to probe material properties and is of fundamental importance for controlling carrier transport.
Nontrivial effects can happen during interface scattering. In geometric optics, it is known that a circularly-polarized light beam undergoes a transverse shift normal to its plane of incidence when reflected at an optical interface, called the Imbert-Fedorov (IF) shift~\cite{Fedorov1955,Imbert1972,Onoda2004,Bliokh2006,Hosten2008,Yin2013}.
Recently, analogous effect has been discovered for electronic systems, showing that transverse shifts also appear for electrons in so-called Weyl semimetals~\cite{Jiang2015,Yang2015b,JiangPRB2016,LWang2017}. In both cases, the spin-orbit coupling (SOC) plays the key role.
The light helicity, corresponding to the photonic spin state, intrinsically couples with the light propagation in the Maxwell equation~\cite{Bliokh2015}; and the low-energy electrons in Weyl semimetals also possess a strong SOC described by the Weyl equation~\cite{Wan2011,Murakami2007}.
Upon scattering, any change in the particle spin would require a change in the orbital motion due to SOC, resulting in the anomalous spatial shift.

There is an intriguing and unique scattering process occurring at the normal-metal-superconductor (NS) interface---the Andreev reflection, in which an incident electron from the normal metal is reflected back as a hole, accompanied by the transfer of a Cooper pair into the superconductor~\cite{Andreev1964,Gennes1966}. Most recently, we find that the transverse shift can also exist in Andreev reflection, if the interface is formed by a spin-orbit-coupled metal and a conventional $s$-wave superconductor~\cite{Liu2017}.
It is important to note that the essential ingredient there is still the assumed strong SOC of the scattered carrier---the shift vanishes when SOC is negligible; whereas the superconductivity only plays a \emph{passive} role as a channel for electron-hole conversion.

Unconventional pairing brings new physics into the picture. By breaking more symmetries than the $U(1)$ gauge symmetry, unconventional pair potentials necessarily have a strong wave-vector dependence~\cite{Sigrist2005}.
Surprisingly, we find that in Andreev reflection from an unconventional superconductor, a sizable transverse shift exists even \emph{in the absence of} SOC, resulting solely from the unconventional pair potential. We show that the unconventional pairing provides an \emph{effective} coupling between the orbital motion and the pseudospin of the electron-hole (Nambu) space, which underlies this exotic effect. Remarkably, the value of the shift is sensitive to the structure of pair potential and manifests characteristic features for each pairing type, as summarized in Table~\ref{T1}. The effect can be detected through electric measurement,
providing a promising new technique for probing the structure of unconventional pairings.

\begin{figure}[t]
\includegraphics[width=8.5cm]{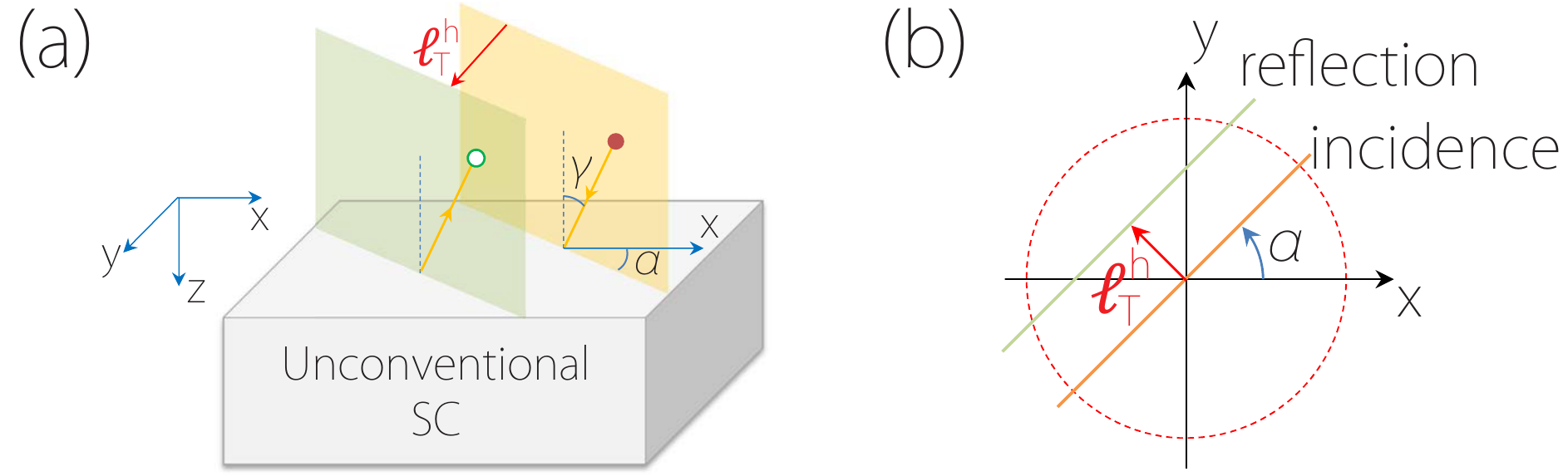}
\caption{
(a) Schematic of the NS junction set-up.  In Andreev reflection, the reflection plane (green-colored) is shifted by distance $\ell_T^h$ from the incident plane (orange-colored) along its normal direction  ($\hat{n}$), due to unconventional pairing in S.  Here, $\hat{n}$ is specified to be along $\hat{k}\times \hat{z}$ with $\hat{k}$ the incident direction.
(b) Top view of the $x$-$y$ plane in (a). For certain pairings, there may also be a finite shift $\ell_T^e$ for normally-reflected electrons (not shown here).}\label{fig_junct}
\end{figure}

{\color{blue}{\em Model.}} Since our goal is to demonstrate the existence of the finite transverse shift, we take a simplest model for a three-dimensional NS junction with a flat interface in the clean limit.  In this work, we focus on the case where the interface is perpendicular to the principle rotation axis (along $z$-direction) of the superconductor [Fig.~\ref{fig_junct}(b)]. Configurations with other interface orientations can be similarly studied. To highlight the role of unconventional pairing, we neglect SOC in the model. Then for each pairing considered in Table~\ref{T1}, the essential physics of scattering at the NS interface (located at $z=0$) can be captured by the Bogoliubov-de Gennes (BdG) equation~\cite{Gennes1966,Blonder1982,Kashiwaya2000} in the following reduced form
\begin{equation}
\left[\begin{array}{cc}
H_{0}-E_{F}+V(z) & \bm{\Delta}(z)\\
\bm{\Delta}^{*}(z) & E_{F}-H_{0}-V(z)
\end{array}\right]\psi=\varepsilon\psi.
\label{BdG}
\end{equation}
Here, $\psi$ is the two-component spinor wave-function in the Nambu space (the real-spin labels are suppressed), $E_F$ is the Fermi energy, and $V(z)=U\Theta(z)+h\delta(z)$ with $U$ the band bottom shift, $h$ the interface barrier potential, and $\Theta$ the Heaviside step function. We take the single-particle Hamiltonian $H_0=-\frac{1}{2m}\nabla^2$ for the normal-metal (N) side ($z<0$), and $H_0=-\frac{1}{2m_\|}(\partial_x^2+\partial_y^2)-\frac{1}{2m_z}\partial_z^2$ for the superconductor (S) side ($z>0$). The difference in the effective masses $m_\|$ and $m_z$ describes the possible uniaxial anisotropy in S. For certain layered superconductors (like cuprates), the Fermi surface is highly anisotropic and may take a cylinder-like shape in the normal state.
Such case can be described using a lattice model, and we find that the essential features of our results remain the same~\cite{SuppMater}.
For concrete calculation, we take the usual step function model for the pair potential $\bm\Delta(z)=\bm\Delta\Theta(z)$~\cite{Blonder1982,Jong1995}. It is a good approximation to the full self-consistent solution for the BdG equation near an interface~\cite{Plehn1991,Hara1993,Plehn1994}, and as we show below, for certain cases with an emergent symmetry, the transverse shift actually does not depend on the detailed $z$-variation of $\bm\Delta(z)$. We consider the weak coupling limit with $E_F-U\gg |\bm \Delta|,\varepsilon$ in the S region, so that the wave-vector for $\bm\Delta$'s $k$-dependence is fixed on the (normal-state) Fermi surface of S, and $\bm\Delta$ only depends on the direction of the wave-vector $\bm k$~\cite{Tanaka1995}.

{\color{blue}{\em Intuitive picture \& symmetry argument.}}  The spatial shift is defined for a confined electron beam undergoing reflection at the interface. The incident geometry is characterized by two angles $\gamma$ and $\alpha$, as illustrated in Fig.~\ref{fig_junct}. The beam is usually modeled by a wave-packet $\bm\Psi$~\cite{Beenakker2009,Jiang2015,Liu2017}, which is assumed to be confined in both real and momentum spaces. The detailed form of $\bm\Psi$ does not concern us for now.

Let's first consider the case when $\bm\Delta$ is of chiral $p$-wave pairing, with $\bm \Delta=\Delta_0 e^{i\chi\phi_k}$, because it offers an intuitive understanding of the physical picture.
Here $\chi=\pm 1$ denotes the chirality, $(\theta_k,\phi_k)$ are the spherical angles of $\bm k$, and $\Delta_0$ is assumed to be independent of $\phi_k$ but may still depend on $\theta_k$. A key observation is that the BdG equation possesses an emergent symmetry:
\begin{equation}
[\hat{\mathcal{H}}_\text{BdG},\hat{\mathcal{J}}_z]=0,
\end{equation}
where $\hat{\mathcal{H}}_\text{BdG}$ is the BdG Hamiltonian in Eq.~(\ref{BdG}) with $\bm \Delta$ taking the chiral $p$-wave form, and
\begin{equation}
\hat{\mathcal{J}}_z=(\hat{\bm r}\times\hat{\bm k})-\frac{1}{2}\chi \hat\tau_z
\end{equation}
resembles an effective angular momentum operator with $\hat\tau_z$ the Pauli matrix corresponding to the Nambu pseudospin-$1/2$. Consequently, the expectation value
$J_z=\langle\bm\Psi|\hat{\mathcal{J}}_z|\bm\Psi\rangle$ evaluated for the beam must conserve during scattering.
For electrons and holes, the expectation values of the Nambu pseudospin are opposite: $\langle\hat\tau_{z}\rangle_{e/h} =\pm 1$.
Because the pseudospin flips in Andreev reflection, the conservation of $J_z$ must dictate a transverse shift $\ell_T^h$ to compensate this change (see Fig.~\ref{fig_junct}), leading to
\begin{equation}
\ell_{T}^{h}=-\frac{\chi}{2k_\|}(\langle\hat\tau_{z}\rangle_h-\langle\hat\tau_{z}\rangle_e)=\frac{\chi}{k_{\parallel}}, \label{chiralpAndr}
\end{equation}
where $k_\|=k_F^N\sin\gamma$, $k_F^N$ is the Fermi wave-vector in N, and $\gamma$ is the incident angle.

This remarkable result demonstrates several points. First, the physical picture becomes clear: the shift here is entirely due to the unconventional pairing which plays the role of an \emph{effective} SOC that couples $k$ and $\tau$. However, the spin here is not the real spin but the Nambu pseudospin, which is \emph{intrinsic} and \emph{unique} for superconductors. The change in pseudospin during Andreev reflection then naturally results in the shift in real space. Second, as resulting from a symmetry argument, Eq.~(\ref{chiralpAndr}) is quite general: as long as the symmetry is preserved, factors like the variation of $\bm \Delta(z)$, the excitation energy, or the interfacial barrier will not affect $\ell_T^h$. Third, the result (\ref{chiralpAndr}) also applies for chiral pairings with higher orbital moments ($|\chi|>1$), such as $d+id$ or $f+if$ pairings. Finally, the shift becomes pronounced when $k_F^N$ (or $\gamma$) is small, e.g., when N side is of doped semiconductor or semimetal with small Fermi surfaces, a feature similar to analogous effects discussed before~\cite{Jiang2015,Yang2015b}.

{\color{blue}{\em Scattering approach.}}  For general cases without a conserved $\hat{\mathcal{J}}_z$, the shift can be obtained via the quantum scattering approach~\cite{Beenakker2009,Jiang2015,Liu2017}, which has been standard in studying the shift in optical and electronic contexts. In this approach, $\bm \Psi$ is expanded using the scattering eigenstates of the system. For example, the incident beam $\bm\Psi^{e+}_{\bm k}(\bm r)=\int d\bm k'\ w(\bm k'-\bm k)\psi^{e+}_{\bm k'}(\bm r)$, where $w$ is the profile of the beam peaked at $\bm k$, and $\psi^{e+}$ is the incident electron eigenstate. At the interface, each partial wave $\psi^{e+}$ gets scattered. Particularly, for Andreev reflection, $\psi^{e+}$ is reflected as $\psi^{h-}$ with amplitude $r_h$, hence the reflected hole beam is given by $\bm\Psi^{h-}_{\bm k}(\bm r)=\int d\bm k'\ w(\bm k'-\bm k)r_h(\bm k')\psi^{h-}_{\bm k'}(\bm r)$. The spatial shift is obtained by comparing the center positions of the two beams at the interface. One easily finds that the shift here takes a simple form
\begin{equation}
\delta \ell_i^{h}=-\frac{\partial}{\partial k_i}\arg(r_{h})\Big|_{\bm k_\|},
\label{shift}
\end{equation}
where $i\in\{x,y\}$, $\bm k_\|$ is the average transverse wave-vector which is conserved in scattering, and $\arg(r_{h})$ is the phase of $r_h$. In this approach, the spatial shift appears as a result of interference between the scattered partial waves. As evident from Eq.~(\ref{shift}), it only depends on (the $k$-dependence of) the phase of the scattering amplitude, not the magnitude.

We have some additional remarks. (i) There may also exists a shift $\delta \ell_i^{e}$ for the normally-reflected electron beam, whose expression takes the same form as Eq.~(\ref{shift}) by replacing $r_h$ with $r_e$. (ii) Although the shift does not depend on the magnitudes of $r_h$ ($r_e$), the intensity of the reflected beam is proportional to $|r_h|^2$ ($|r_e|^2$). (iii) The shift has both longitudinal (analogous to the Goos-H\"{a}nchen shift in optics~\cite{Goos1947}) and transverse components with reference to the incident plane. In this work, we focus on the transverse shift: $\ell_T^{e(h)}\equiv \delta\bm\ell^{e(h)}\cdot \hat n$, where $\hat n$ is the normal direction of the incident plane, as illustrated in Fig.~\ref{fig_junct}. (iv) The scattering approach is quite general. Unlike the semiclassical approach which requires the quasiparticle wavelength to be small compared with the perturbation length scale~\cite{Yang2015b}, the scattering approach does not suffer from such constraint and applies for sharp interfaces and/or large wavelengths as well~\cite{Beenakker2009,Jiang2015,Liu2017}.

This approach is applied to study the transverse shift for each type of the pair potentials. The calculation is straightforward, and the key results are tabulated in Table~\ref{T1}. For conventional $s$-wave pairing, one easily checks that $\ell_T^{e}=\ell_T^{h}=0$, consistent with our previous findings~\cite{Liu2017}. In contrast, the transverse shift can become sizable if S side is of unconventional pairing, and it possesses features tied with the pairing symmetry.
We discuss two cases below.

{\renewcommand{\arraystretch}{1.2}
{\begin{table}[tb]
\newcommand{\tabincell}[2]{\begin{tabular}{@{}#1@{}}#2\end{tabular}}
\centering
\begin{tabular}{c|c|c|c|c|c|c}
 \hline\hline
 \multicolumn{2}{c|}{\multirow{2}{*}{\tabincell{c} {Pair \\ potential}}}&
 \multicolumn{2}{c|}{Expression}&
  \multirow{2}{*}{\tabincell{c}{Period \\ in $\alpha$}} &
  \multirow{2}{*}{\tabincell{c} {vanish for  \\ $\varepsilon>|\bm{\Delta}(\alpha)|$}} &
 \multirow{2}{*}{\tabincell{c}{No. of  \\ SZ}}
  \\
 \cline{3-4}
    \multicolumn{2}{c|}{} & $\ell_{T}^{e}$ & $\ell_{T}^{h}$ &  {} & {}&{}\\
 \hline
 Chiral & $\Delta_0 e^{i\chi \phi_k}$ & 0 & $\frac{\chi}{k_F^{N}\sin \gamma}$ & $-$ &No & 0 \\
 \hline
 $p_x$ & $\Delta_0 \cos \phi_k$ & \multicolumn{2}{c|}{\multirow{4}{*}{$\ell_T^e \approx  \ell_T^h $}} & $\pi$ &\multirow{4}{*}{Yes} & 2 \\
  \cline{1-2}  \cline{5-5} \cline{7-7}
  $p_y$& $\Delta_0 \sin \phi_k$ & \multicolumn{2}{c|}{} & $\pi$ & {} & 2 \\
 \cline{1-2}  \cline{5-5} \cline{7-7}
  $d_{x^2-y^2}$& $\Delta_0 \cos 2\phi_k$ & \multicolumn{2}{c|}{} & $\pi/2$ & {} & 4 \\
  \cline{1-2}  \cline{5-5} \cline{7-7}
  $d_{xy}$& $\Delta_0 \sin 2\phi_k$ & \multicolumn{2}{c|}{} & $\pi/2$ & {} & 4 \\
 \hline\hline
 \end{tabular}
\caption{Features of the transverse shift for typical unconventional pair potentials. ``No. of SZ" stands for the number of suppressed zones when the rotation angle $\alpha$ varies from 0 to $2\pi$. }\label{T1}
\end{table}}
}

\begin{figure}[t]
\includegraphics[width=8.5cm]{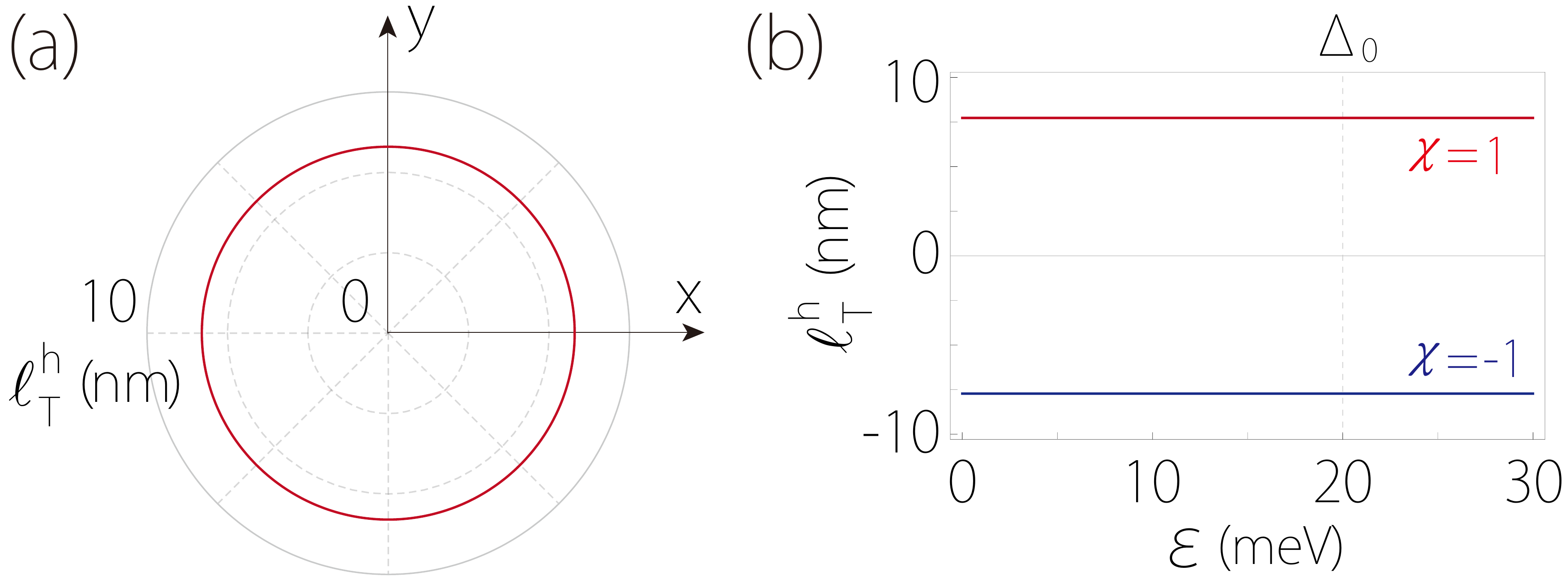}
\caption{
Transverse shift $\ell_T^{h}$ for chiral $p$-wave pairing versus (a) rotation angle $\alpha$  (here $\chi=+1$), and (b) the excitation energy $\varepsilon$.
$\ell_T^{h}$  is independent of $\varepsilon$ and $\Delta_0$, and its sign depends on $\chi$.
{Here, we take  $\Delta_0=20$ meV, $\gamma=\pi/12$, $E_F=0.1$ eV, and $m=0.1 m_e$.} }\label{fig_pwave}
\end{figure}

{\color{blue}{\em Chiral $p$-wave.}} Let's re-visit the case for chiral $p$-wave pair potential using the scattering approach. Straightforward calculation shows that $\arg(r_h)=-\chi\phi_k$ and $\arg(r_e)$ is independent of $k$ ~\cite{SuppMater}. Then according to Eq.~(\ref{shift}),
\begin{equation}\label{chiralp}
\ell_T^h=\chi/k_\|,\qquad \ell_T^{e}=0,
\end{equation}
which exactly recovers the result we have obtained using the symmetry argument. This result is independent of the incident angles, the excitation energy, and the
parameters for S side such as the pairing gap $\Delta_0$, except that its sign depends on the chirality $\chi$ (see Fig.~\ref{fig_pwave}).

\begin{figure}[t]
\includegraphics[width=8.5cm]{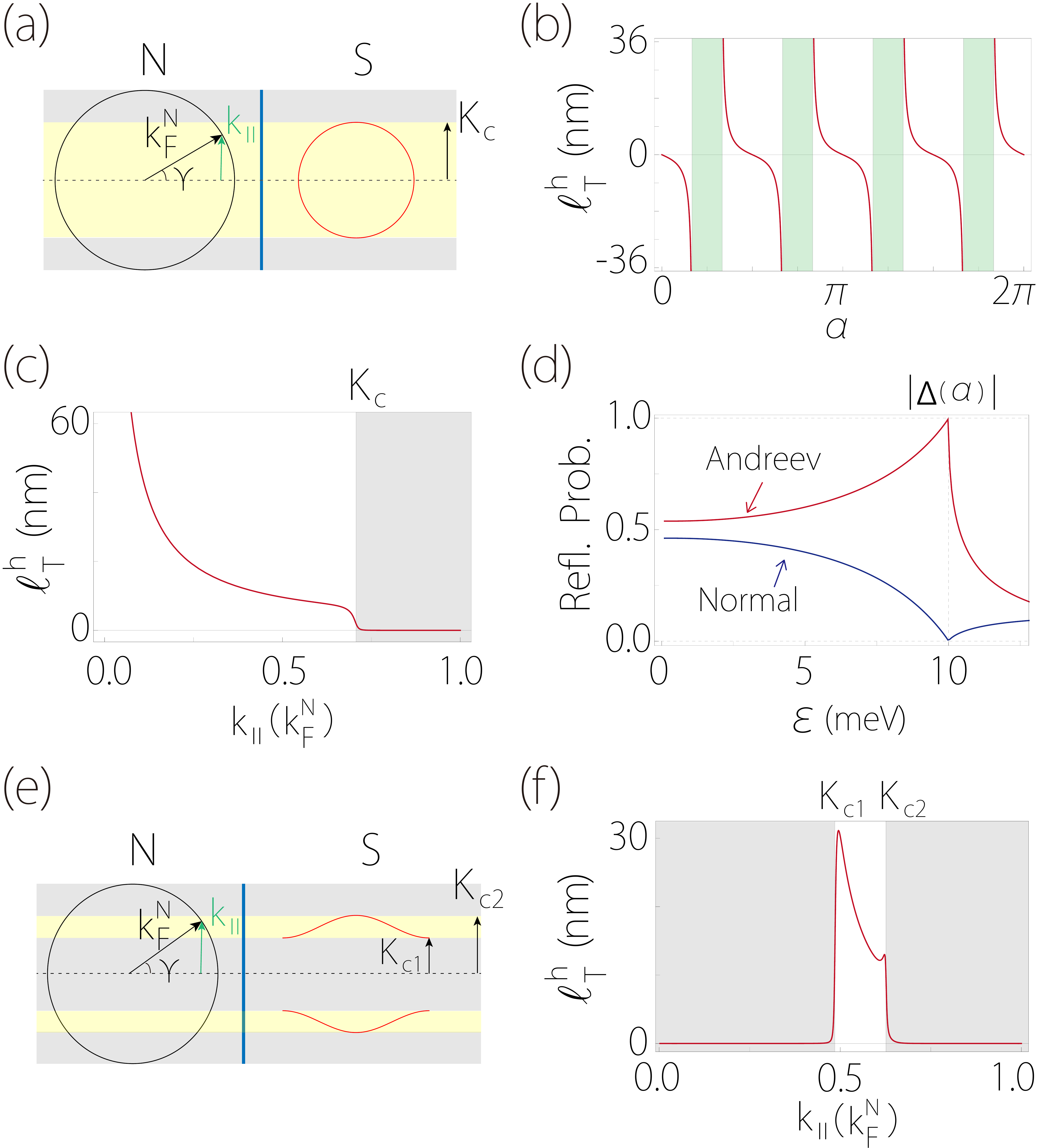}
\caption{
Results for $d_{x^2-y^2}$-wave pairing. (a-d) are for the S side with an ellipsoidal Fermi surface, and (e-f) are for S with a cylinder-like Fermi surface.
(a) Schematic figure showing the Fermi surfaces of N and S. $K_c$ denotes the maximum magnitude of transverse wave-vector on the S Fermi surface.
(b) $\ell_T^{h}$ versus  $\alpha$. The green shaded regions indicate the suppressed zones, in which  $\varepsilon>|\bm \Delta(\alpha)|$.
(c) $\ell_T^{h}$ versus $k_\|$.
Corresponding to (a), $\ell_T^{h}$ is suppressed  when $k_\|>K_c$, as denoted by the gray shaded region.
(d) Reflection probabilities versus $\varepsilon$ for normal and Andreev reflections.
(e) illustrates the case when the S Fermi surface is of open cylinder-like shape. $K_{c1}$ and $K_{c2}$ denote the lower and upper bounds for the transverse wave-vector on the S Fermi surface.
For such case, the qualitative features in (b) and (d) remain the same~\cite{SuppMater}. The main difference is that the shift is now suppressed in regions except for $K_{c1}<k_\|<K_{c2}$, as shown in (f).
In (a-d),
we take $\Delta_0=20\ \rm{meV}$, $E_F=0.4\  \rm{eV}$, $U=0.2\  \rm{eV}$, $h=0.3\ \rm{eV}\cdot \rm{nm}$, and  $m_{\parallel}=m_z=m=0.1 m_e$.
We set $\varepsilon=10\ \rm{meV}$ and  {$\gamma=\pi/12$ }  in (b); { $\alpha=-\pi/6$ and  $\varepsilon=8\ \rm{meV}$  in (c); $\alpha=\pi/6$ and $\gamma=\pi/5$ in (d)}.
The parameters for (f) are presented in Supplemental Material~\cite{SuppMater}.}\label{fig_dwave}
\end{figure}

{\color{blue}{\em $d_{x^2-y^2}$-wave.}} As another example, we consider the $d_{x^2-y^2}$-wave pairing, with $\bm \Delta=\Delta_0\cos(2\phi_k)$. Our calculation gives that~\cite{SuppMater}
\begin{equation}\label{dwave}
\ell_{T}^h \propto \sin(4\alpha)\Theta(|\Delta_{0}\cos2\alpha|-\varepsilon),
\end{equation}
and $\ell_T^e\approx\ell_T^h$. The expression in Eq.~(\ref{dwave}) highlights the dependence on the rotation angle $\alpha$ and the excitation energy $\varepsilon$. The results are plotted in Fig.~\ref{fig_dwave}.

From Eq.~(\ref{dwave}) and Fig.~\ref{fig_dwave}, we observe the following key features for the shift. (i) $\ell_T^{e(h)}$ has a period of $\pi/2$ in $\alpha$, and it flips sign at multiples of $\pi/4$ [Fig.~\ref{fig_dwave}(b)]. (ii) $\ell_T^{e(h)}$ is sensitive to the gap magnitude. As indicated by the step function in Eq.~(\ref{dwave}), it is suppressed for excitation energies above the pairing gap at the incident wave-vector. (iii) Particularly, due to the nodal structure of the gap, for a fixed excitation energy $\varepsilon$, there must appear multiple zones in $\alpha$ where $\ell_T^{e(h)}$ is suppressed [see Fig.~\ref{fig_dwave}(b)]. The center of each suppressed zone coincides with a node. (iv) $\ell_T^{e(h)}$ is also suppressed when $k_\|$ is away from the Fermi surface of S side, as indicated in Fig.~\ref{fig_dwave}(c) and \ref{fig_dwave}(f), where we compare the results for a closed ellipsoidal Fermi surface and for an open cylinder-like Fermi surface. This can be understood by noticing that the effect of pair potential diminishes away from the Fermi surface.

The above features of the shifts encode rich information about the unconventional gap structure, including the $d$-wave symmetry [Feature (i)], the gap magnitude profile [Feature (ii)], and the node position [Feature (iii)]. Thus, by detecting the effect, one can extract important information about the unconventional superconductor. Feature (iv) also offers information on the geometry of the Fermi surface.

{\color{blue}{\em Discussion.}} Here, we have revealed a fundamentally new effect---the anomalous shifts in Andreev reflection generated by unconventional pairings. Distinct from the previous works~\cite{Jiang2015,Yang2015b,Liu2017}, where the shift invariably originates from the SOC in the N region and vanishes if the SOC is negligible,
the shift here is purely from the unconventional pairing in the S region, and it exists without the need of SOC. Because of this fundamental difference, the shifts here manifest the characteristics of unconventional pairings in S, such as the highly anisotropic behavior with respect to the incident direction as in Fig.~\ref{fig_dwave}(b), which is tied to the anisotropic character of the $d_{x^2-y^2}$-pairing; whereas the shifts in Ref.~\onlinecite{Liu2017} instead reflect the SOC pattern of the N region.


\begin{figure}[htbp]
\includegraphics[width=8.8cm]{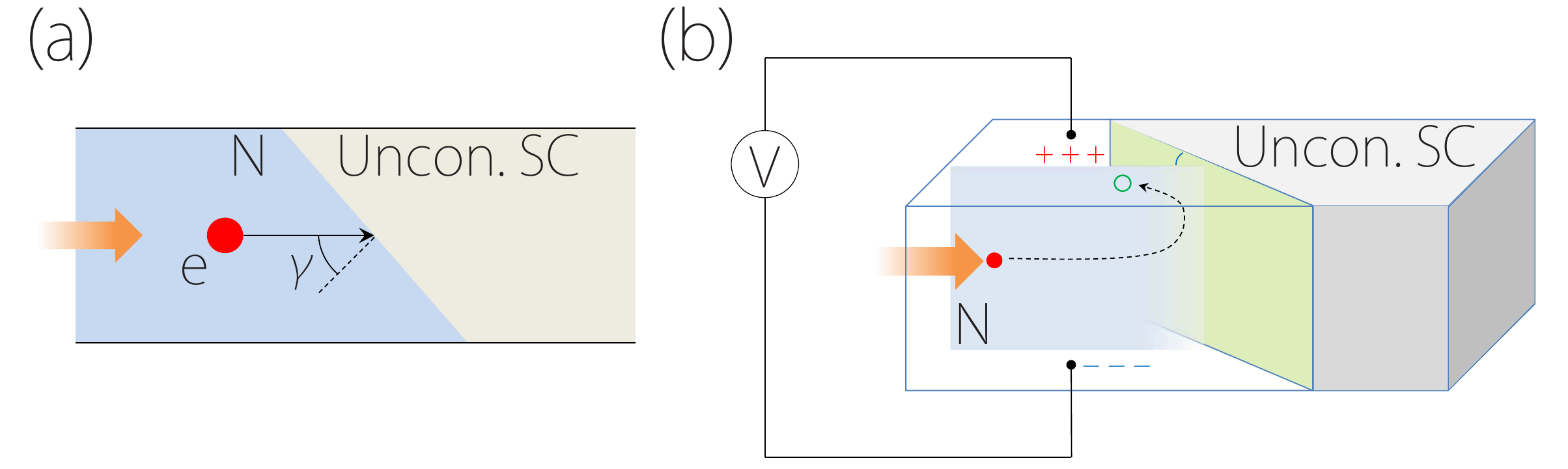}
\caption{
Schematic (a) top view and (b) side view of a possible NS junction geometry for experimental detection. Electrons are driven to the interface with a finite average incident angle. The effect of transverse shift (in $z$-direction) induces a net surface charge accumulation near the junction on the N side, which can be detected as a voltage difference between top and bottom surfaces. Here we illustrate the case when Andreev reflections dominate the interface scattering.}\label{fig_dect}
\end{figure}

To experimentally probe the effect, we suggest a simple NS junction geometry as illustrated in Fig.~\ref{fig_dect}, in which the electrons are driven towards the interface with a finite average incident angle. The transverse shift then leads to the surface charge accumulation as indicated in Fig.~\ref{fig_dect}(b), which can be detected as a voltage signal between top and bottom surfaces.  With an order of magnitude estimation~\cite{SuppMater}, we find that the voltage signal can be on the order of mV, readily detectable in experiment. We also suggest a second setup to amplify the shift through multiple scattering in an SNS structure, which leads to a large anomalous velocity up to $10^4$ m/s~\cite{SuppMater}.
With more delicate setups, e.g., by using local gates and collimators similar to those in electron optics~\cite{Spector1990,Dragoman1999}, one could control the angles $(\gamma,\alpha)$ of the incident beam, and the excitation energy $\varepsilon$ can be controlled by the junction bias voltage. Then by mapping out the signal dependence on $(\gamma,\alpha,\varepsilon)$, one can extract the features of the shifts and in principle characterize the gap structure for unconventional superconductors.

Here it should be noted that: While for chiral pairings, the voltage signal is solely due to the shifts in Andreev reflection; for nonchiral pairings (like $d_{x^2-y^2}$-wave), the signal may have contributions from both normal and Andreev reflections. Since $\ell_T^e\approx\ell_T^h$, the net result depends on the probabilities ($|r_h|^2$ vs $|r_e|^2$) of the two processes. There could be interesting competition between the two when tuning the excitation energy. Generally, for $\varepsilon$ close to the superconducting gap, $|r_h|^2$ would dominate over $|r_e|^2$~\cite{Blonder1982}, so in this case the signal would be dominated by the shifts in Andreev reflection.

We remark that real unconventional superconductor materials could have other complicated features, such as multiple Fermi surfaces, multiple bands with different pairing magnitudes, and possible surface/interface bound states~\cite{Sigrist2005,Tsuei2000,Mackenzie2003,Beiranvand2016,Kallin2016}. How these features would affect the anomalous shifts are interesting questions to explore in future studies. Nevertheless, our analysis suggests that a nonzero shift (hence the resulting voltage) is generally expected, owing to the coupling between the Nambu pseudospin and the orbital motion as generated by the unconventional pair potential. Although its detailed profile requires more accurate material-specific modeling, the key features for the shift, like the period in $\alpha$ and the gap dependence as listed in Table~\ref{T1}, should be robust, since they are determined by the overall characteristic associated with the symmetry of unconventional pairings.  This also helps to distinguish the signal from the shift against random noises such as from the impurities or interface roughness.
Finally, when SOC effect is included, it can generate an additional contribution to the shift in Andreev reflection~\cite{Liu2017}. However, its dependence on the incident geometry and the excitation energy will be distinct from that due to the unconventional pairings.

\bibliography{Tshift_UCSC_ref}

\begin{thebibliography}{44}%
\makeatletter
\providecommand \@ifxundefined [1]{%
 \@ifx{#1\undefined}
}%
\providecommand \@ifnum [1]{%
 \ifnum #1\expandafter \@firstoftwo
 \else \expandafter \@secondoftwo
 \fi
}%
\providecommand \@ifx [1]{%
 \ifx #1\expandafter \@firstoftwo
 \else \expandafter \@secondoftwo
 \fi
}%
\providecommand \natexlab [1]{#1}%
\providecommand \enquote  [1]{``#1''}%
\providecommand \bibnamefont  [1]{#1}%
\providecommand \bibfnamefont [1]{#1}%
\providecommand \citenamefont [1]{#1}%
\providecommand \href@noop [0]{\@secondoftwo}%
\providecommand \href [0]{\begingroup \@sanitize@url \@href}%
\providecommand \@href[1]{\@@startlink{#1}\@@href}%
\providecommand \@@href[1]{\endgroup#1\@@endlink}%
\providecommand \@sanitize@url [0]{\catcode `\\12\catcode `\$12\catcode
  `\&12\catcode `\#12\catcode `\^12\catcode `\_12\catcode `\%12\relax}%
\providecommand \@@startlink[1]{}%
\providecommand \@@endlink[0]{}%
\providecommand \url  [0]{\begingroup\@sanitize@url \@url }%
\providecommand \@url [1]{\endgroup\@href {#1}{\urlprefix }}%
\providecommand \urlprefix  [0]{URL }%
\providecommand \Eprint [0]{\href }%
\providecommand \doibase [0]{http://dx.doi.org/}%
\providecommand \selectlanguage [0]{\@gobble}%
\providecommand \bibinfo  [0]{\@secondoftwo}%
\providecommand \bibfield  [0]{\@secondoftwo}%
\providecommand \translation [1]{[#1]}%
\providecommand \BibitemOpen [0]{}%
\providecommand \bibitemStop [0]{}%
\providecommand \bibitemNoStop [0]{.\EOS\space}%
\providecommand \EOS [0]{\spacefactor3000\relax}%
\providecommand \BibitemShut  [1]{\csname bibitem#1\endcsname}%
\let\auto@bib@innerbib\@empty
\bibitem [{\citenamefont {Fedorov}(1955)}]{Fedorov1955}%
  \BibitemOpen
  \bibfield  {author} {\bibinfo {author} {\bibfnamefont {F.}~\bibnamefont
  {Fedorov}},\ }\href@noop {} {\bibfield  {journal} {\bibinfo  {journal} {Dokl.
  Akad. Nauk SSSR}\ }\textbf {\bibinfo {volume} {105}},\ \bibinfo {pages} {465}
  (\bibinfo {year} {1955})}\BibitemShut {NoStop}%
\bibitem [{\citenamefont {Imbert}(1972)}]{Imbert1972}%
  \BibitemOpen
  \bibfield  {author} {\bibinfo {author} {\bibfnamefont {C.}~\bibnamefont
  {Imbert}},\ }\href@noop {} {\bibfield  {journal} {\bibinfo  {journal} {Phys.
  Rev. D}\ }\textbf {\bibinfo {volume} {5}},\ \bibinfo {pages} {787} (\bibinfo
  {year} {1972})}\BibitemShut {NoStop}%
\bibitem [{\citenamefont {Onoda}\ \emph {et~al.}(2004)\citenamefont {Onoda},
  \citenamefont {Murakami},\ and\ \citenamefont {Nagaosa}}]{Onoda2004}%
  \BibitemOpen
  \bibfield  {author} {\bibinfo {author} {\bibfnamefont {M.}~\bibnamefont
  {Onoda}}, \bibinfo {author} {\bibfnamefont {S.}~\bibnamefont {Murakami}}, \
  and\ \bibinfo {author} {\bibfnamefont {N.}~\bibnamefont {Nagaosa}},\
  }\href@noop {} {\bibfield  {journal} {\bibinfo  {journal} {Phys. Rev. Lett.}\
  }\textbf {\bibinfo {volume} {93}},\ \bibinfo {pages} {083901} (\bibinfo
  {year} {2004})}\BibitemShut {NoStop}%
\bibitem [{\citenamefont {Bliokh}\ and\ \citenamefont
  {Bliokh}(2006)}]{Bliokh2006}%
  \BibitemOpen
  \bibfield  {author} {\bibinfo {author} {\bibfnamefont {K.~Y.}\ \bibnamefont
  {Bliokh}}\ and\ \bibinfo {author} {\bibfnamefont {Y.~P.}\ \bibnamefont
  {Bliokh}},\ }\href@noop {} {\bibfield  {journal} {\bibinfo  {journal} {Phys.
  Rev. Lett.}\ }\textbf {\bibinfo {volume} {96}},\ \bibinfo {pages} {073903}
  (\bibinfo {year} {2006})}\BibitemShut {NoStop}%
\bibitem [{\citenamefont {Hosten}\ and\ \citenamefont
  {Kwiat}(2008)}]{Hosten2008}%
  \BibitemOpen
  \bibfield  {author} {\bibinfo {author} {\bibfnamefont {O.}~\bibnamefont
  {Hosten}}\ and\ \bibinfo {author} {\bibfnamefont {P.}~\bibnamefont {Kwiat}},\
  }\href@noop {} {\bibfield  {journal} {\bibinfo  {journal} {Science}\ }\textbf
  {\bibinfo {volume} {319}},\ \bibinfo {pages} {787} (\bibinfo {year}
  {2008})}\BibitemShut {NoStop}%
\bibitem [{\citenamefont {Yin}\ \emph {et~al.}(2013)\citenamefont {Yin},
  \citenamefont {Ye}, \citenamefont {Rho}, \citenamefont {Wang},\ and\
  \citenamefont {Zhang}}]{Yin2013}%
  \BibitemOpen
  \bibfield  {author} {\bibinfo {author} {\bibfnamefont {X.}~\bibnamefont
  {Yin}}, \bibinfo {author} {\bibfnamefont {Z.}~\bibnamefont {Ye}}, \bibinfo
  {author} {\bibfnamefont {J.}~\bibnamefont {Rho}}, \bibinfo {author}
  {\bibfnamefont {Y.}~\bibnamefont {Wang}}, \ and\ \bibinfo {author}
  {\bibfnamefont {X.}~\bibnamefont {Zhang}},\ }\href@noop {} {\bibfield
  {journal} {\bibinfo  {journal} {Science}\ }\textbf {\bibinfo {volume}
  {339}},\ \bibinfo {pages} {1405} (\bibinfo {year} {2013})}\BibitemShut
  {NoStop}%
\bibitem [{\citenamefont {Jiang}\ \emph {et~al.}(2015)\citenamefont {Jiang},
  \citenamefont {Jiang}, \citenamefont {Liu}, \citenamefont {Sun},\ and\
  \citenamefont {Xie}}]{Jiang2015}%
  \BibitemOpen
  \bibfield  {author} {\bibinfo {author} {\bibfnamefont {Q.-D.}\ \bibnamefont
  {Jiang}}, \bibinfo {author} {\bibfnamefont {H.}~\bibnamefont {Jiang}},
  \bibinfo {author} {\bibfnamefont {H.}~\bibnamefont {Liu}}, \bibinfo {author}
  {\bibfnamefont {Q.-F.}\ \bibnamefont {Sun}}, \ and\ \bibinfo {author}
  {\bibfnamefont {X.~C.}\ \bibnamefont {Xie}},\ }\href@noop {} {\bibfield
  {journal} {\bibinfo  {journal} {Phys. Rev. Lett.}\ }\textbf {\bibinfo
  {volume} {115}},\ \bibinfo {pages} {156602} (\bibinfo {year}
  {2015})}\BibitemShut {NoStop}%
\bibitem [{\citenamefont {Yang}\ \emph {et~al.}(2015)\citenamefont {Yang},
  \citenamefont {Pan},\ and\ \citenamefont {Zhang}}]{Yang2015b}%
  \BibitemOpen
  \bibfield  {author} {\bibinfo {author} {\bibfnamefont {S.~A.}\ \bibnamefont
  {Yang}}, \bibinfo {author} {\bibfnamefont {H.}~\bibnamefont {Pan}}, \ and\
  \bibinfo {author} {\bibfnamefont {F.}~\bibnamefont {Zhang}},\ }\href@noop {}
  {\bibfield  {journal} {\bibinfo  {journal} {Phys. Rev. Lett.}\ }\textbf
  {\bibinfo {volume} {115}},\ \bibinfo {pages} {156603} (\bibinfo {year}
  {2015})}\BibitemShut {NoStop}%
\bibitem [{\citenamefont {Jiang}\ \emph {et~al.}(2016)\citenamefont {Jiang},
  \citenamefont {Jiang}, \citenamefont {Liu}, \citenamefont {Sun},\ and\
  \citenamefont {Xie}}]{JiangPRB2016}%
  \BibitemOpen
  \bibfield  {author} {\bibinfo {author} {\bibfnamefont {Q.-D.}\ \bibnamefont
  {Jiang}}, \bibinfo {author} {\bibfnamefont {H.}~\bibnamefont {Jiang}},
  \bibinfo {author} {\bibfnamefont {H.}~\bibnamefont {Liu}}, \bibinfo {author}
  {\bibfnamefont {Q.-F.}\ \bibnamefont {Sun}}, \ and\ \bibinfo {author}
  {\bibfnamefont {X.~C.}\ \bibnamefont {Xie}},\ }\href@noop {} {\bibfield
  {journal} {\bibinfo  {journal} {Phys. Rev. B}\ }\textbf {\bibinfo {volume}
  {93}},\ \bibinfo {pages} {195165} (\bibinfo {year} {2016})}\BibitemShut
  {NoStop}%
\bibitem [{LWa()}]{LWang2017}%
  \BibitemOpen
  \href@noop {} {}\bibinfo {note} {L. Wang and S.-K. Jian,
  arXiv:1703.03168.}\BibitemShut {Stop}%
\bibitem [{\citenamefont {Bliokh}\ \emph {et~al.}(2015)\citenamefont {Bliokh},
  \citenamefont {Rodriguez-Fortuno}, \citenamefont {Nori},\ and\ \citenamefont
  {Zayats}}]{Bliokh2015}%
  \BibitemOpen
  \bibfield  {author} {\bibinfo {author} {\bibfnamefont {K.~Y.}\ \bibnamefont
  {Bliokh}}, \bibinfo {author} {\bibfnamefont {F.~J.}\ \bibnamefont
  {Rodriguez-Fortuno}}, \bibinfo {author} {\bibfnamefont {F.}~\bibnamefont
  {Nori}}, \ and\ \bibinfo {author} {\bibfnamefont {A.~V.}\ \bibnamefont
  {Zayats}},\ }\href@noop {} {\bibfield  {journal} {\bibinfo  {journal} {Nat
  Photon}\ }\textbf {\bibinfo {volume} {9}},\ \bibinfo {pages} {796} (\bibinfo
  {year} {2015})}\BibitemShut {NoStop}%
\bibitem [{\citenamefont {Wan}\ \emph {et~al.}(2011)\citenamefont {Wan},
  \citenamefont {Turner}, \citenamefont {Vishwanath},\ and\ \citenamefont
  {Savrasov}}]{Wan2011}%
  \BibitemOpen
  \bibfield  {author} {\bibinfo {author} {\bibfnamefont {X.}~\bibnamefont
  {Wan}}, \bibinfo {author} {\bibfnamefont {A.~M.}\ \bibnamefont {Turner}},
  \bibinfo {author} {\bibfnamefont {A.}~\bibnamefont {Vishwanath}}, \ and\
  \bibinfo {author} {\bibfnamefont {S.~Y.}\ \bibnamefont {Savrasov}},\
  }\href@noop {} {\bibfield  {journal} {\bibinfo  {journal} {Phys. Rev. B}\
  }\textbf {\bibinfo {volume} {83}},\ \bibinfo {pages} {205101} (\bibinfo
  {year} {2011})}\BibitemShut {NoStop}%
\bibitem [{\citenamefont {Murakami}(2007)}]{Murakami2007}%
  \BibitemOpen
  \bibfield  {author} {\bibinfo {author} {\bibfnamefont {S.}~\bibnamefont
  {Murakami}},\ }\href@noop {} {\bibfield  {journal} {\bibinfo  {journal} {New
  Journal of Physics}\ }\textbf {\bibinfo {volume} {9}},\ \bibinfo {pages}
  {356} (\bibinfo {year} {2007})}\BibitemShut {NoStop}%
\bibitem [{\citenamefont {Andreev}(1964)}]{Andreev1964}%
  \BibitemOpen
  \bibfield  {author} {\bibinfo {author} {\bibfnamefont {A.~F.}\ \bibnamefont
  {Andreev}},\ }\href@noop {} {\bibfield  {journal} {\bibinfo  {journal} {Sov.
  Phys. JETP}\ }\textbf {\bibinfo {volume} {19}},\ \bibinfo {pages} {1228}
  (\bibinfo {year} {1964})}\BibitemShut {NoStop}%
\bibitem [{\citenamefont {de~Gennes}(1966)}]{Gennes1966}%
  \BibitemOpen
  \bibfield  {author} {\bibinfo {author} {\bibfnamefont {P.~G.}\ \bibnamefont
  {de~Gennes}},\ }\href@noop {} {\emph {\bibinfo {title} {Superconductivity in
  Metals and Alloys}}}\ (\bibinfo  {publisher} {Benjamin},\ \bibinfo {address}
  {New York},\ \bibinfo {year} {1966})\BibitemShut {NoStop}%
\bibitem [{\citenamefont {Liu}\ \emph {et~al.}(2017)\citenamefont {Liu},
  \citenamefont {Yu},\ and\ \citenamefont {Yang}}]{Liu2017}%
  \BibitemOpen
  \bibfield  {author} {\bibinfo {author} {\bibfnamefont {Y.}~\bibnamefont
  {Liu}}, \bibinfo {author} {\bibfnamefont {Z.-M.}\ \bibnamefont {Yu}}, \ and\
  \bibinfo {author} {\bibfnamefont {S.~A.}\ \bibnamefont {Yang}},\ }\href@noop
  {} {\bibfield  {journal} {\bibinfo  {journal} {Phys. Rev. B}\ }\textbf
  {\bibinfo {volume} {96}},\ \bibinfo {pages} {121101} (\bibinfo {year}
  {2017})}\BibitemShut {NoStop}%
\bibitem [{\citenamefont {Sigrist}\ \emph {et~al.}(2005)\citenamefont
  {Sigrist}, \citenamefont {Avella},\ and\ \citenamefont
  {Mancini}}]{Sigrist2005}%
  \BibitemOpen
  \bibfield  {author} {\bibinfo {author} {\bibfnamefont {M.}~\bibnamefont
  {Sigrist}}, \bibinfo {author} {\bibfnamefont {A.}~\bibnamefont {Avella}}, \
  and\ \bibinfo {author} {\bibfnamefont {F.}~\bibnamefont {Mancini}},\
  }\href@noop {} {\bibfield  {journal} {\bibinfo  {journal} {AIP Conference
  Proceedings}\ }\textbf {\bibinfo {volume} {789}},\ \bibinfo {pages} {165}
  (\bibinfo {year} {2005})}\BibitemShut {NoStop}%
\bibitem [{\citenamefont {Blonder}\ \emph {et~al.}(1982)\citenamefont
  {Blonder}, \citenamefont {Tinkham},\ and\ \citenamefont
  {Klapwijk}}]{Blonder1982}%
  \BibitemOpen
  \bibfield  {author} {\bibinfo {author} {\bibfnamefont {G.~E.}\ \bibnamefont
  {Blonder}}, \bibinfo {author} {\bibfnamefont {M.}~\bibnamefont {Tinkham}}, \
  and\ \bibinfo {author} {\bibfnamefont {T.~M.}\ \bibnamefont {Klapwijk}},\
  }\href@noop {} {\bibfield  {journal} {\bibinfo  {journal} {Phys. Rev. B}\
  }\textbf {\bibinfo {volume} {25}},\ \bibinfo {pages} {4515} (\bibinfo {year}
  {1982})}\BibitemShut {NoStop}%
\bibitem [{\citenamefont {Kashiwaya}\ and\ \citenamefont
  {Tanaka}(2000)}]{Kashiwaya2000}%
  \BibitemOpen
  \bibfield  {author} {\bibinfo {author} {\bibfnamefont {S.}~\bibnamefont
  {Kashiwaya}}\ and\ \bibinfo {author} {\bibfnamefont {Y.}~\bibnamefont
  {Tanaka}},\ }\href@noop {} {\bibfield  {journal} {\bibinfo  {journal}
  {Reports on Progress in Physics}\ }\textbf {\bibinfo {volume} {63}},\
  \bibinfo {pages} {1641} (\bibinfo {year} {2000})}\BibitemShut {NoStop}%
\bibitem [{Sup()}]{SuppMater}%
  \BibitemOpen
  \href@noop {} {}\bibinfo {note} {See Supplemental Material for detailed
  calculations of scattering amplitudes and transverse shift for reflected
  wave-packet, and calculations in the lattice model, which includes Refs.
  \cite{Blonder1982,Kashiwaya2000,Beiranvand2016,
  BenDaniel1966,Beenakker2009,Jiang2015,su2015form,Ando1991,Khomyakov2005,sotoodeh2000,Murzin2001,Ohno1999,murakami2003,ashcroft2005,gall2016,bauer2002}.}\BibitemShut
  {Stop}%
\bibitem [{\citenamefont {de~Jong}\ and\ \citenamefont
  {Beenakker}(1995)}]{Jong1995}%
  \BibitemOpen
  \bibfield  {author} {\bibinfo {author} {\bibfnamefont {M.~J.~M.}\
  \bibnamefont {de~Jong}}\ and\ \bibinfo {author} {\bibfnamefont {C.~W.~J.}\
  \bibnamefont {Beenakker}},\ }\href@noop {} {\bibfield  {journal} {\bibinfo
  {journal} {Phys. Rev. Lett.}\ }\textbf {\bibinfo {volume} {74}},\ \bibinfo
  {pages} {1657} (\bibinfo {year} {1995})}\BibitemShut {NoStop}%
\bibitem [{\citenamefont {Plehn}\ \emph {et~al.}(1991)\citenamefont {Plehn},
  \citenamefont {Gunsenheimer},\ and\ \citenamefont {Kümmel}}]{Plehn1991}%
  \BibitemOpen
  \bibfield  {author} {\bibinfo {author} {\bibfnamefont {H.}~\bibnamefont
  {Plehn}}, \bibinfo {author} {\bibfnamefont {U.}~\bibnamefont {Gunsenheimer}},
  \ and\ \bibinfo {author} {\bibfnamefont {R.}~\bibnamefont {Kümmel}},\
  }\href@noop {} {\bibfield  {journal} {\bibinfo  {journal} {Journal of Low
  Temperature Physics}\ }\textbf {\bibinfo {volume} {83}},\ \bibinfo {pages}
  {71} (\bibinfo {year} {1991})}\BibitemShut {NoStop}%
\bibitem [{\citenamefont {Hara}\ \emph {et~al.}(1993)\citenamefont {Hara},
  \citenamefont {Ashida},\ and\ \citenamefont {Nagai}}]{Hara1993}%
  \BibitemOpen
  \bibfield  {author} {\bibinfo {author} {\bibfnamefont {J.}~\bibnamefont
  {Hara}}, \bibinfo {author} {\bibfnamefont {M.}~\bibnamefont {Ashida}}, \ and\
  \bibinfo {author} {\bibfnamefont {K.}~\bibnamefont {Nagai}},\ }\href@noop {}
  {\bibfield  {journal} {\bibinfo  {journal} {Phys. Rev. B}\ }\textbf {\bibinfo
  {volume} {47}},\ \bibinfo {pages} {11263} (\bibinfo {year}
  {1993})}\BibitemShut {NoStop}%
\bibitem [{\citenamefont {Plehn}\ \emph {et~al.}(1994)\citenamefont {Plehn},
  \citenamefont {Wacker},\ and\ \citenamefont {K\"ummel}}]{Plehn1994}%
  \BibitemOpen
  \bibfield  {author} {\bibinfo {author} {\bibfnamefont {H.}~\bibnamefont
  {Plehn}}, \bibinfo {author} {\bibfnamefont {O.-J.}\ \bibnamefont {Wacker}}, \
  and\ \bibinfo {author} {\bibfnamefont {R.}~\bibnamefont {K\"ummel}},\
  }\href@noop {} {\bibfield  {journal} {\bibinfo  {journal} {Phys. Rev. B}\
  }\textbf {\bibinfo {volume} {49}},\ \bibinfo {pages} {12140} (\bibinfo {year}
  {1994})}\BibitemShut {NoStop}%
\bibitem [{\citenamefont {Tanaka}\ and\ \citenamefont
  {Kashiwaya}(1995)}]{Tanaka1995}%
  \BibitemOpen
  \bibfield  {author} {\bibinfo {author} {\bibfnamefont {Y.}~\bibnamefont
  {Tanaka}}\ and\ \bibinfo {author} {\bibfnamefont {S.}~\bibnamefont
  {Kashiwaya}},\ }\href@noop {} {\bibfield  {journal} {\bibinfo  {journal}
  {Phys. Rev. Lett.}\ }\textbf {\bibinfo {volume} {74}},\ \bibinfo {pages}
  {3451} (\bibinfo {year} {1995})}\BibitemShut {NoStop}%
\bibitem [{\citenamefont {Beenakker}\ \emph {et~al.}(2009)\citenamefont
  {Beenakker}, \citenamefont {Sepkhanov}, \citenamefont {Akhmerov},\ and\
  \citenamefont {Tworzyd\l{}o}}]{Beenakker2009}%
  \BibitemOpen
  \bibfield  {author} {\bibinfo {author} {\bibfnamefont {C.~W.~J.}\
  \bibnamefont {Beenakker}}, \bibinfo {author} {\bibfnamefont {R.~A.}\
  \bibnamefont {Sepkhanov}}, \bibinfo {author} {\bibfnamefont {A.~R.}\
  \bibnamefont {Akhmerov}}, \ and\ \bibinfo {author} {\bibfnamefont
  {J.}~\bibnamefont {Tworzyd\l{}o}},\ }\href@noop {} {\bibfield  {journal}
  {\bibinfo  {journal} {Phys. Rev. Lett.}\ }\textbf {\bibinfo {volume} {102}},\
  \bibinfo {pages} {146804} (\bibinfo {year} {2009})}\BibitemShut {NoStop}%
\bibitem [{\citenamefont {Goos}\ and\ \citenamefont
  {H\"{a}nchen}(1947)}]{Goos1947}%
  \BibitemOpen
  \bibfield  {author} {\bibinfo {author} {\bibfnamefont {F.}~\bibnamefont
  {Goos}}\ and\ \bibinfo {author} {\bibfnamefont {H.}~\bibnamefont
  {H\"{a}nchen}},\ }\href@noop {} {\bibfield  {journal} {\bibinfo  {journal}
  {Ann. Phys.}\ }\textbf {\bibinfo {volume} {436}},\ \bibinfo {pages} {333}
  (\bibinfo {year} {1947})}\BibitemShut {NoStop}%
\bibitem [{\citenamefont {Spector}\ \emph {et~al.}(1990)\citenamefont
  {Spector}, \citenamefont {Stormer}, \citenamefont {Baldwin}, \citenamefont
  {Pfeiffer},\ and\ \citenamefont {West}}]{Spector1990}%
  \BibitemOpen
  \bibfield  {author} {\bibinfo {author} {\bibfnamefont {J.}~\bibnamefont
  {Spector}}, \bibinfo {author} {\bibfnamefont {H.~L.}\ \bibnamefont
  {Stormer}}, \bibinfo {author} {\bibfnamefont {K.~W.}\ \bibnamefont
  {Baldwin}}, \bibinfo {author} {\bibfnamefont {L.~N.}\ \bibnamefont
  {Pfeiffer}}, \ and\ \bibinfo {author} {\bibfnamefont {K.~W.}\ \bibnamefont
  {West}},\ }\href@noop {} {\bibfield  {journal} {\bibinfo  {journal} {Appl.
  Phys. Lett.}\ }\textbf {\bibinfo {volume} {56}},\ \bibinfo {pages} {1290}
  (\bibinfo {year} {1990})}\BibitemShut {NoStop}%
\bibitem [{\citenamefont {Dragoman}\ and\ \citenamefont
  {Dragoman}(1999)}]{Dragoman1999}%
  \BibitemOpen
  \bibfield  {author} {\bibinfo {author} {\bibfnamefont {D.}~\bibnamefont
  {Dragoman}}\ and\ \bibinfo {author} {\bibfnamefont {M.}~\bibnamefont
  {Dragoman}},\ }\href@noop {} {\bibfield  {journal} {\bibinfo  {journal}
  {Progress in Quantum Electronics}\ }\textbf {\bibinfo {volume} {23}},\
  \bibinfo {pages} {131} (\bibinfo {year} {1999})}\BibitemShut {NoStop}%
\bibitem [{\citenamefont {Tsuei}\ and\ \citenamefont
  {Kirtley}(2000)}]{Tsuei2000}%
  \BibitemOpen
  \bibfield  {author} {\bibinfo {author} {\bibfnamefont {C.~C.}\ \bibnamefont
  {Tsuei}}\ and\ \bibinfo {author} {\bibfnamefont {J.~R.}\ \bibnamefont
  {Kirtley}},\ }\href@noop {} {\bibfield  {journal} {\bibinfo  {journal} {Rev.
  Mod. Phys.}\ }\textbf {\bibinfo {volume} {72}},\ \bibinfo {pages} {969}
  (\bibinfo {year} {2000})}\BibitemShut {NoStop}%
\bibitem [{\citenamefont {Mackenzie}\ and\ \citenamefont
  {Maeno}(2003)}]{Mackenzie2003}%
  \BibitemOpen
  \bibfield  {author} {\bibinfo {author} {\bibfnamefont {A.~P.}\ \bibnamefont
  {Mackenzie}}\ and\ \bibinfo {author} {\bibfnamefont {Y.}~\bibnamefont
  {Maeno}},\ }\href@noop {} {\bibfield  {journal} {\bibinfo  {journal} {Rev.
  Mod. Phys.}\ }\textbf {\bibinfo {volume} {75}},\ \bibinfo {pages} {657}
  (\bibinfo {year} {2003})}\BibitemShut {NoStop}%
\bibitem [{\citenamefont {Beiranvand}\ \emph {et~al.}(2016)\citenamefont
  {Beiranvand}, \citenamefont {Hamzehpour},\ and\ \citenamefont
  {Alidoust}}]{Beiranvand2016}%
  \BibitemOpen
  \bibfield  {author} {\bibinfo {author} {\bibfnamefont {R.}~\bibnamefont
  {Beiranvand}}, \bibinfo {author} {\bibfnamefont {H.}~\bibnamefont
  {Hamzehpour}}, \ and\ \bibinfo {author} {\bibfnamefont {M.}~\bibnamefont
  {Alidoust}},\ }\href@noop {} {\bibfield  {journal} {\bibinfo  {journal}
  {Phys. Rev. B}\ }\textbf {\bibinfo {volume} {94}},\ \bibinfo {pages} {125415}
  (\bibinfo {year} {2016})}\BibitemShut {NoStop}%
\bibitem [{\citenamefont {Kallin}\ and\ \citenamefont
  {Berlinsky}(2016)}]{Kallin2016}%
  \BibitemOpen
  \bibfield  {author} {\bibinfo {author} {\bibfnamefont {C.}~\bibnamefont
  {Kallin}}\ and\ \bibinfo {author} {\bibfnamefont {J.}~\bibnamefont
  {Berlinsky}},\ }\href@noop {} {\bibfield  {journal} {\bibinfo  {journal}
  {Reports on Progress in Physics}\ }\textbf {\bibinfo {volume} {79}},\
  \bibinfo {pages} {054502} (\bibinfo {year} {2016})}\BibitemShut {NoStop}%
\bibitem [{\citenamefont {BenDaniel}\ and\ \citenamefont
  {Duke}(1966)}]{BenDaniel1966}%
  \BibitemOpen
  \bibfield  {author} {\bibinfo {author} {\bibfnamefont {D.~J.}\ \bibnamefont
  {BenDaniel}}\ and\ \bibinfo {author} {\bibfnamefont {C.~B.}\ \bibnamefont
  {Duke}},\ }\href@noop {} {\bibfield  {journal} {\bibinfo  {journal} {Phys.
  Rev.}\ }\textbf {\bibinfo {volume} {152}},\ \bibinfo {pages} {683} (\bibinfo
  {year} {1966})}\BibitemShut {NoStop}%
\bibitem [{\citenamefont {Su}\ \emph {et~al.}(2015)\citenamefont {Su},
  \citenamefont {Liao},\ and\ \citenamefont {Li}}]{su2015form}%
  \BibitemOpen
  \bibfield  {author} {\bibinfo {author} {\bibfnamefont {Y.}~\bibnamefont
  {Su}}, \bibinfo {author} {\bibfnamefont {H.}~\bibnamefont {Liao}}, \ and\
  \bibinfo {author} {\bibfnamefont {T.}~\bibnamefont {Li}},\ }\href@noop {}
  {\bibfield  {journal} {\bibinfo  {journal} {Journal of Physics: Condensed
  Matter}\ }\textbf {\bibinfo {volume} {27}},\ \bibinfo {pages} {105702}
  (\bibinfo {year} {2015})}\BibitemShut {NoStop}%
\bibitem [{\citenamefont {Ando}(1991)}]{Ando1991}%
  \BibitemOpen
  \bibfield  {author} {\bibinfo {author} {\bibfnamefont {T.}~\bibnamefont
  {Ando}},\ }\href@noop {} {\bibfield  {journal} {\bibinfo  {journal} {Phys.
  Rev. B}\ }\textbf {\bibinfo {volume} {44}},\ \bibinfo {pages} {8017}
  (\bibinfo {year} {1991})}\BibitemShut {NoStop}%
\bibitem [{\citenamefont {Khomyakov}\ \emph {et~al.}(2005)\citenamefont
  {Khomyakov}, \citenamefont {Brocks}, \citenamefont {Karpan}, \citenamefont
  {Zwierzycki},\ and\ \citenamefont {Kelly}}]{Khomyakov2005}%
  \BibitemOpen
  \bibfield  {author} {\bibinfo {author} {\bibfnamefont {P.~A.}\ \bibnamefont
  {Khomyakov}}, \bibinfo {author} {\bibfnamefont {G.}~\bibnamefont {Brocks}},
  \bibinfo {author} {\bibfnamefont {V.}~\bibnamefont {Karpan}}, \bibinfo
  {author} {\bibfnamefont {M.}~\bibnamefont {Zwierzycki}}, \ and\ \bibinfo
  {author} {\bibfnamefont {P.~J.}\ \bibnamefont {Kelly}},\ }\href@noop {}
  {\bibfield  {journal} {\bibinfo  {journal} {Phys. Rev. B}\ }\textbf {\bibinfo
  {volume} {72}},\ \bibinfo {pages} {035450} (\bibinfo {year}
  {2005})}\BibitemShut {NoStop}%
\bibitem [{\citenamefont {Sotoodeh}\ \emph {et~al.}(2000)\citenamefont
  {Sotoodeh}, \citenamefont {Khalid},\ and\ \citenamefont
  {Rezazadeh}}]{sotoodeh2000}%
  \BibitemOpen
  \bibfield  {author} {\bibinfo {author} {\bibfnamefont {M.}~\bibnamefont
  {Sotoodeh}}, \bibinfo {author} {\bibfnamefont {A.}~\bibnamefont {Khalid}}, \
  and\ \bibinfo {author} {\bibfnamefont {A.}~\bibnamefont {Rezazadeh}},\
  }\href@noop {} {\bibfield  {journal} {\bibinfo  {journal} {Journal of applied
  physics}\ }\textbf {\bibinfo {volume} {87}},\ \bibinfo {pages} {2890}
  (\bibinfo {year} {2000})}\BibitemShut {NoStop}%
\bibitem [{\citenamefont {Murzin}\ \emph {et~al.}(2001)\citenamefont {Murzin},
  \citenamefont {Weiss}, \citenamefont {Jansen},\ and\ \citenamefont
  {Eberl}}]{Murzin2001}%
  \BibitemOpen
  \bibfield  {author} {\bibinfo {author} {\bibfnamefont {S.~S.}\ \bibnamefont
  {Murzin}}, \bibinfo {author} {\bibfnamefont {M.}~\bibnamefont {Weiss}},
  \bibinfo {author} {\bibfnamefont {A.~G.~M.}\ \bibnamefont {Jansen}}, \ and\
  \bibinfo {author} {\bibfnamefont {K.}~\bibnamefont {Eberl}},\ }\href@noop {}
  {\bibfield  {journal} {\bibinfo  {journal} {Phys. Rev. B}\ }\textbf {\bibinfo
  {volume} {64}},\ \bibinfo {pages} {233309} (\bibinfo {year}
  {2001})}\BibitemShut {NoStop}%
\bibitem [{\citenamefont {Ohno}\ \emph {et~al.}(1999)\citenamefont {Ohno},
  \citenamefont {Terauchi}, \citenamefont {Adachi}, \citenamefont {Matsukura},\
  and\ \citenamefont {Ohno}}]{Ohno1999}%
  \BibitemOpen
  \bibfield  {author} {\bibinfo {author} {\bibfnamefont {Y.}~\bibnamefont
  {Ohno}}, \bibinfo {author} {\bibfnamefont {R.}~\bibnamefont {Terauchi}},
  \bibinfo {author} {\bibfnamefont {T.}~\bibnamefont {Adachi}}, \bibinfo
  {author} {\bibfnamefont {F.}~\bibnamefont {Matsukura}}, \ and\ \bibinfo
  {author} {\bibfnamefont {H.}~\bibnamefont {Ohno}},\ }\href@noop {} {\bibfield
   {journal} {\bibinfo  {journal} {Phys. Rev. Lett.}\ }\textbf {\bibinfo
  {volume} {83}},\ \bibinfo {pages} {4196} (\bibinfo {year}
  {1999})}\BibitemShut {NoStop}%
\bibitem [{\citenamefont {Murakami}\ \emph {et~al.}(2003)\citenamefont
  {Murakami}, \citenamefont {Nagaosa},\ and\ \citenamefont
  {Zhang}}]{murakami2003}%
  \BibitemOpen
  \bibfield  {author} {\bibinfo {author} {\bibfnamefont {S.}~\bibnamefont
  {Murakami}}, \bibinfo {author} {\bibfnamefont {N.}~\bibnamefont {Nagaosa}}, \
  and\ \bibinfo {author} {\bibfnamefont {S.-C.}\ \bibnamefont {Zhang}},\
  }\href@noop {} {\bibfield  {journal} {\bibinfo  {journal} {Science}\ }\textbf
  {\bibinfo {volume} {301}},\ \bibinfo {pages} {1348} (\bibinfo {year}
  {2003})}\BibitemShut {NoStop}%
\bibitem [{\citenamefont {Ashcroft}\ and\ \citenamefont
  {Mermin}(1976)}]{ashcroft2005}%
  \BibitemOpen
  \bibfield  {author} {\bibinfo {author} {\bibfnamefont {N.~W.}\ \bibnamefont
  {Ashcroft}}\ and\ \bibinfo {author} {\bibfnamefont {N.~D.}\ \bibnamefont
  {Mermin}},\ }\href@noop {} {\bibfield  {journal} {\bibinfo  {journal} {Solid
  State Physics}\ } (\bibinfo {year} {1976})}\BibitemShut {NoStop}%
\bibitem [{\citenamefont {Gall}(2016)}]{gall2016}%
  \BibitemOpen
  \bibfield  {author} {\bibinfo {author} {\bibfnamefont {D.}~\bibnamefont
  {Gall}},\ }\href@noop {} {\bibfield  {journal} {\bibinfo  {journal} {Journal
  of Applied Physics}\ }\textbf {\bibinfo {volume} {119}},\ \bibinfo {pages}
  {085101} (\bibinfo {year} {2016})}\BibitemShut {NoStop}%
\bibitem [{\citenamefont {Bauer}\ and\ \citenamefont
  {Aeschlimann}(2002)}]{bauer2002}%
  \BibitemOpen
  \bibfield  {author} {\bibinfo {author} {\bibfnamefont {M.}~\bibnamefont
  {Bauer}}\ and\ \bibinfo {author} {\bibfnamefont {M.}~\bibnamefont
  {Aeschlimann}},\ }\href@noop {} {\bibfield  {journal} {\bibinfo  {journal}
  {Journal of electron spectroscopy and related phenomena}\ }\textbf {\bibinfo
  {volume} {124}},\ \bibinfo {pages} {225} (\bibinfo {year}
  {2002})}\BibitemShut {NoStop}%
\end{thebibliography}%


\end{document}


\title{Supplemental Material for ``Unconventional pairing induced transverse shift in Andreev reflection''}

\author{Zhi-Ming Yu}
\affiliation{Research Laboratory for Quantum Materials, Singapore University of Technology and Design, Singapore 487372, Singapore}

\author{Ying Liu}
\affiliation{Research Laboratory for Quantum Materials, Singapore University of Technology and Design, Singapore 487372, Singapore}

\author{Yugui Yao}
\affiliation{Beijing Key Laboratory of Nanophotonics and Ultrafine Optoelectronic Systems, School of Physics, Beijing Institute of Technology, Beijing 100081, China}

\author{Shengyuan A. Yang}
\affiliation{Research Laboratory for Quantum Materials, Singapore University of Technology and Design, Singapore 487372, Singapore}



\maketitle
\maketitle
\section{Calculation of scattering amplitudes}
We use the quantum scattering approach to obtain the scattering amplitudes at the NS interface. This approach has been well established in studying the transport through NS junctions with both conventional and unconventional pair potentials~\cite{BTK,Kashiwaya,Beiranvand}. For our model
[Eq.~(1) in the main text], the scattering state for an incident electron from the N side can be written as
\begin{eqnarray}
\psi(\boldsymbol{r}) & = & \begin{cases}
\psi^{e+}+r_{e}\psi^{e-}+r_{h}\psi^{h-}, & z<0,\\
t_{+}\psi_{+}^{\text{S}}+t_{-}\psi_{-}^{\text{S}}, & z>0,
\end{cases}\label{eq:psi}
\end{eqnarray}
where $r_{e}$ ($r_{h}$) is the normal (Andreev) reflection amplitude,
and $t_{\pm}$ are the transmission amplitudes. The basis states on the N side ($z<0$) are given by
\begin{eqnarray}
\psi^{e+}=e^{ik_{x}x+ik_{y}y+ik_{z}^{e}z}\left[\begin{array}{c}
1\\
0
\end{array}\right], & \psi^{e-}=e^{ik_{x}x+ik_{y}y-ik_{z}^{e}z}\left[\begin{array}{c}
1\\
0
\end{array}\right], & \psi^{h-}=e^{ik_{x}x+ik_{y}y+ik_{z}^{h}z}\left[\begin{array}{c}
0\\
1
\end{array}\right],\label{basis}
\end{eqnarray}
and the basis states on the S side are obtained as
\begin{equation}
\psi_{\pm}^{\text{S}}=e^{ik_{x}x+ik_{y}y+ik_{\pm}^{\text{S}}z}\left[\begin{array}{c}
1\\
\eta_{\pm}
\end{array}\right],
\end{equation}
where $\eta_{\pm}=\frac{\boldsymbol{\Delta}_{\pm}^{*}}{\varepsilon\pm\sqrt{\varepsilon^{2}-|\boldsymbol{\Delta}_{\pm}|^{2}}}$
with $\boldsymbol{\Delta}_{\pm}=\boldsymbol{\Delta}(k_{x},k_{y},k_{\pm}^{{\rm S}})$.
Under the condition that $E_{F}-U\gg|\boldsymbol{\Delta}|,\varepsilon$,
one has
\begin{equation}
k_{z}^{e/h}=\sqrt{2mE_{F}-k_{\parallel}^{2}},\ \ \ k_{\pm}^{\text{S}}=\pm\sqrt{2m_{z}(E_{F}-U)-m_{z}k_{\parallel}^{2}/m_{\parallel}}\label{eq:ks}
\end{equation}
with $k_{\parallel}=\sqrt{k_{x}^{2}+k_{y}^{2}}$.

To solve the scattering states and evaluate the scattering amplitudes, we need boundary conditions to connect the states at $z=0$. Here, since the effective mass may differ for the two sides of the NS interface, the appropriate single-particle Hamiltonian $H_0$ (in Eq.~(1) of main text) should be written as $H_0=-\sum_{i=x,y,z}\partial_{i}\frac{1}{2m_i(z)}\partial_{i}$~\cite{BDD}, where $m_i(z<0)=m$; $m_{x(y)}(z>0)=m_{\parallel}$, and $m_z(z>0)=m_z$. The correct boundary condition must ensure the quasiparticle current conservation~\cite{BTK}. These requirements lead to the following boundary conditions:
\begin{eqnarray}
\psi|_{z=0^{-}}&=&\psi|_{z=0^{+}},\\
\frac{1}{m}\partial_{z}\psi\big|_{z=0^{-}}&=&\frac{1}{m_{z}}\partial_{z}\psi\big|_{z=0^{+}}-2h\psi(0).\label{eq:BCs}
\end{eqnarray}
The second equation above is analogous to the BenDaniel-Duke boundary condition for semiconductor heterostructures~\cite{BDD}. Here we are most interested in $r_e$ and $r_h$, because they are needed for calculating the transverse shifts in reflections. Straightforward calculations lead to
\begin{eqnarray}\label{amp}
r_{e}=X^{-1}\left[(\eta_{+}-\eta_{-})(1-\Gamma^{2}-4Z^{2}-2iZ)\right], & \  & r_{h}=-4X^{-1}\eta_{+}\eta_{-}\Gamma,
\end{eqnarray}
with $Z=\frac{mh}{k_{z}^{e}}$, $\Gamma=\frac{mk_{+}^{S}}{m_{z}k_{z}^{e}}$,
$\zeta=\frac{1+\Gamma^{2}+4Z^{2}}{2\Gamma}$, and $X=2\Gamma[\eta_{+}(\zeta-1)-\eta_{-}(\zeta+1)]$.


\section{Calculation of the anomalous shift.}

The spatial shift is defined for the trajectory of a quasiparticle beam, which is usually modeled using a wave-packet~\cite{Jiang,Beenakker}. Consider the wave-packet of an incident electron built from the plane-wave basis states:
\begin{eqnarray}
\bm \Psi^{e+}_{\bm k}(\boldsymbol{r}) & = & \int d\bm{k}'\ w(\bm{k}'-\bm{k})\psi^{e+}_{\bm k'}(\bm r),
\end{eqnarray}
where the function $w$ is the beam profile required to make it spatially confined. Note
that none of the results will depend on the specific form of $w$~\cite{Jiang,Beenakker}.
In calculation, for definiteness, one usually chooses $w$ to have
a Gaussian form: $w(\bm{q})=\prod_{i}w_{i}(q_{i})$
with $w_{i}(q_{i})=(\sqrt{2\pi}W_{i})^{-1}\exp[-q_i^{2}/(2W_{i}^{2})]$,
with a width $W_{i}$ for the $i$-th component (which is small compared with the Fermi wave-vector). Then the wave-packet
for the reflected electron (hole) is given by $\bm \Psi^{e-(h-)}(\boldsymbol{r})=\int d\bm{k}'\ w(\bm{k}'-\bm{k})r_{e(h)}\psi^{e-(h-)}_{\bm k'}(\bm{r})$. The forms of the basis states have been presented in Eq.~(\ref{basis}).

To analyze the spatial shift, we compare the wave-packet center positions at the interface $(z=0)$~\cite{Jiang,Beenakker}. For example, consider the Andreev-reflected hole wave-packet. By expanding the phase of the amplitude $r_h$ to first order around $\bm k_\|=(k_x, k_y)$, the $k$-integral of the wave-packet gives that
$
\bm \Psi^{h-} \propto  \prod_{i=x,y} e^{-W_{i}^{2} (r_i+\frac{\partial}{\partial k_i}\arg (r_{h})|_{\bm{k}_\|})^2/2}.
$
Comparing with the incident electron wave-packet
$
\bm \Psi^{e+} \propto  \prod_{i=x,y} e^{-W_{i}^{2} r_i^2/2},
$
one finds that the reflected hole has a relative spatial shift of
$\delta \ell^h_i= -\frac{\partial}{\partial k_i}\arg (r_{h})\big|_{\bm{k}_\|}$ ($i=x,y$). The shift for the reflected electron can be obtained in a similar way.
These are the results presented in Eq.~(2) of the main text.

\section{Transverse shifts for the two example pairings.}
First, for the chiral $p$-wave pairing, we have $\boldsymbol{\Delta}_{+}=\boldsymbol{\Delta}_{-}=\Delta_{0}e^{i\chi\phi_{k}}$, and
$\eta_{+}\eta_{-}=e^{-2i\chi\phi_{k}}$. According to Eq.~(\ref{amp}), we find that
\begin{eqnarray}
r_{e}=\frac{(\eta_{+}^{\prime}-\eta_{-}^{\prime})(1-\Gamma^{2}-4Z^{2}-2iZ)}{2\Gamma[\eta_{+}^{\prime}(\zeta-1)-\eta_{-}^{\prime}(\zeta+1)]}, & \  & r_{h}=\frac{-2}{\eta_{+}^{\prime}(\zeta-1)-\eta_{-}^{\prime}(\zeta+1)}e^{-i\chi\phi_{k}},
\end{eqnarray}
with $\eta_{\pm}^{\prime}=\frac{\Delta_{0}}{\varepsilon\pm\sqrt{\varepsilon^{2}-\Delta_{0}^{2}}}$.
Straightforward calculations with Eq.~(2) lead to the result in Eq.~(3) of the main text:
$
\ell_{T}^{e}=0$ and $\ell_{T}^{h}=\chi/k_{\|}.
$
As we have stated in the main text, this result of $\ell_T^h$ can be obtained from a symmetry argument, and it is independent of the parameters on the S side except for the chirality $\chi$.

Second, for the $d_{x^{2}-y^{2}}$-wave pairing,
we have $\boldsymbol{\Delta}_{+}=\boldsymbol{\Delta}_{-}=\Delta_{0}\cos(2\phi_{k})$,
and $\eta_{\pm}=e^{\pm i\vartheta}$ with $\vartheta=i\cosh^{-1}|\frac{\varepsilon}{\bm{\Delta}}|$
for $|\varepsilon|>|\bm{\Delta}|$, and $\vartheta=-\cos^{-1}\frac{\varepsilon}{\bm{\Delta}}$
for $|\varepsilon|<|\bm{\Delta}|$. Then the reflection amplitudes are obtained as
\begin{eqnarray}
r_{e}=-\frac{\sin\vartheta[2Z+i(1-\Gamma^{2}-4Z^{2})]}{2\Gamma}r_{h}, & \ \  & r_{h}=\frac{1}{\cos\vartheta-i\zeta\sin\vartheta}.
\end{eqnarray}
The shifts are found by plugging these expressions into Eq.~(2) of  the main text. We find that $\ell_{T}^{e}\approx\ell_{T}^{h}$ under the weak coupling condition that $E_{F}-U\gg|\boldsymbol{\Delta}|,\varepsilon$ (their difference is of terms at least on the order of $(|\Delta|,\varepsilon)/(E_F-U))$. We find that
\begin{equation}
\ell_{T}^{e}\approx\ell_{T}^{h}=-\frac{\rho\zeta\Delta_{0}^{2}}{k_{\parallel}\varepsilon^{2}(1+\zeta^{2}/\rho^{2})}\sin(4\alpha)\Theta(|\Delta_{0}\cos2\alpha|-\varepsilon),
\end{equation}
with $\rho=\varepsilon/\sqrt{\varepsilon^{2}-\Delta_{0}^{2}\cos^{2}(2\alpha)}$. We note that the shifts vanish when $\varepsilon>|\bm{\Delta}(\alpha)|=|\Delta_{0}\cos2\alpha|$,
because the phases of the scattering amplitudes become independent of the transverse wave-vectors in this case.
Furthermore, we find that $\ell_{T}^{e(h)}$ is also suppressed when $k_\parallel$ is away from the Fermi surface of the S side.

\begin{figure}[hbp]
\includegraphics[width=14cm]{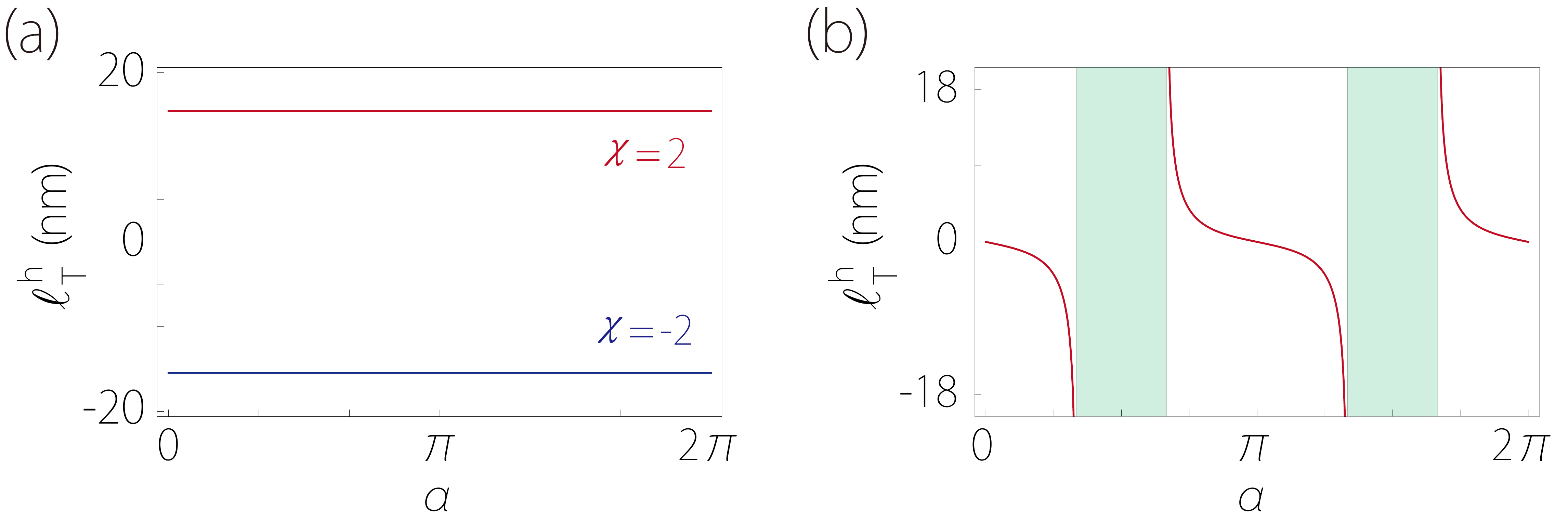}
\caption{Transverse shift for chiral $d$-wave pairing and $p_{x}$-type pairing.
(a), $\ell_T^h$ versus $\alpha$ for chiral $d$-wave pairing ($\bm \Delta=\Delta_0 e^{i\chi\phi_k}$) with $\chi=\pm2$.
(b), $\ell_T^h$ versus $\alpha$ for $p_{x}$-type pairing ($\bm \Delta=\Delta_0\cos\phi_k$). The green shaded regions in (b) are the suppressed zones.
We set $\gamma=\pi/12$, $E_F=0.1$ eV, and $m=0.1 m_e$ in (a); and
$\varepsilon=10\ \rm{meV}$,   $\gamma=\pi/12$,  $\Delta_0=20\ \rm{meV}$, $E_F=0.4\  \rm{eV}$, $U=0.2\  \rm{eV}$, $h=0.3\ \rm{eV}\cdot \rm{nm}$, and  $m_{\parallel}=m_z=m=0.1 m_e$ with $m_e$ the free electron mass  in (b)
\label{fig:S1}}
\end{figure}

Similar analysis can be applied to other types of unconventional pairings,
and their main features are presented
in Table I of the main text. In Fig.~\ref{fig:S1}, we also plot the $\alpha$-dependence of $\ell_{T}^{h}$ for chiral $d$-wave pairing and
$p_{x}$-type pairing.

\section{Calculation in the lattice model}

As mentioned in the main text, for certain layered superconductor like cuperates, the Fermi surface is highly anisotropic and may take a
cylinder-like shape, which is open along the $z$-direction. Such case may be described using a lattice model.

For example, considering the $d_{x^{2}-y^{2}}$-type pairing, a lattice model for the NS junction may be written as \cite{Liao}
\begin{eqnarray}\label{TB}
\hat{H} & = & -\sum_{i,j,l}\left(tc_{i+1,j,l}^{\dagger}c_{i,j,l}+tc_{i,j+1,l}^{\dagger}c_{i,j,l}+t_{l}c_{i,j,l+1}^{\dagger}c_{i,j,l}+{\rm H.c.}\right)-\sum_{i,j,l}(E_{F}-V_{i,j,l})c_{i,j,l}^{\dagger}c_{i,j,l}\nonumber \\
 &  & -\sum_{l>0,\delta=\pm1}\sum_{i,j}\Delta\left(c_{i,j,l}^{\dagger}c_{i+\delta,j,l}^{\dagger}-c_{i,j,l}^{\dagger}c_{i,j+\delta,l}^{\dagger}+{\rm H.c.}\right),
\end{eqnarray}
where $c_{i,j,l}^{\dagger}(c_{i,j,l})$ is the creation (annihilation) operator for an electron on the site labeled by ($i,j,l$) in a cubic lattice, $t$ and $t_l$  are the hopping parameters. Let the NS interface be at $l=0$. We set $t_{l}=t$ for
$l<0$; $t_{l}=t_{h}$ for $l=0$; and $t_{l}=t^{\prime}$ for $l>0$.

Assuming the junction is extended along $x$- and $y$-direction, the transverse wave-vector $\bm k_\|=(k_x,k_y)$ is conserved during scattering. We can focus on a particular scattering channel with a fixed $\bm k_\|$ by doing Fourier transforms along $x$ and $y$-directions for model (\ref{TB}). This leads to an effectively one directional model for each $\bm k_\|$:
\begin{eqnarray}\label{kTB}
\hat{H} & = & -\sum_{l,\bm k_{\parallel}=(k_{x},k_{y})}\left(t\cos k_{x}+t\cos k_{y}+E_{F}-V_{l}\right)\left(c_{l,\bm k_{\parallel}}^{\dagger}c_{l,\bm k_{\parallel}}-c_{l,\bm k_{\parallel}}c_{l,\bm k_{\parallel}}^{\dagger}\right)\nonumber \\
 &  & \qquad -\sum_{l,\bm k_{\parallel}}\frac{t_{l}}{2}\left(c_{l+1,\bm k_{\parallel}}^{\dagger}c_{l,\bm k_{\parallel}}+c_{l,\bm k_{\parallel}}^{\dagger}c_{l+1,\bm k_{\parallel}}-c_{l,\bm k_{\parallel}}c_{l+1,\bm k_{\parallel}}^{\dagger}-c_{l+1,\bm k_{\parallel}}c_{l,\bm k_{\parallel}}^{\dagger}\right)\nonumber \\
 &  & \qquad -2\Delta\sum_{l>0,\bm k_{\parallel}}\left(\cos k_{x}-\cos k_{y}\right)\left(c_{l,\bm k_{\parallel}}^{\dagger}c_{l,-\bm k_{\parallel}}^{\dagger}+c_{l,-\bm k_{\parallel}}c_{l,\bm k_{\parallel}}\right).
\end{eqnarray}

Using the Nambu basis $\alpha_{l}=(\begin{array}{cc}
c_{l,\bm k_{\parallel}}, & c_{l,-\bm k_{\parallel}}^{\dagger}\end{array})^{T}$, we can put the BdG Hamiltonian into a block tridiagonal form
\begin{equation}
\hat{H}=\sum_{l,l^{\prime}}\alpha_{l}^{\dagger}{\rm \bm{H}}_{l,l^{\prime}}\alpha_{l^{\prime}},
\end{equation}
where the elements vanish for $|l-l'|>1$. The intra-layer elements ${\rm \bm{H}}_{l,l}$ for model (\ref{kTB}) is
\begin{equation}\label{S20}
{\rm \bm{H}}_{l,l}=\left[\begin{array}{cc}
-(t\cos k_{x}+t\cos k_{y}+E_{F}-V_{l}) & -2\Delta(\cos k_{x}-\cos k_{y})\Theta(l)\\
-2\Delta(\cos k_{x}-\cos k_{y})\Theta(l) & t\cos k_{x}+t\cos k_{y}+E_{F}-V_{l}
\end{array}\right],
\end{equation}
and the inter-layer elements are
\begin{equation}
{\rm \bm{H}}_{l+1,l}=\left[\begin{array}{cc}
-t_{l}/2 & 0\\
0 & t_{l}/2
\end{array}\right],\ \ \ {\rm \bm{H}}_{l,l+1}={\rm \bm{H}}_{l+1,l}^{\dagger}.
\end{equation}

We note that since the transverse wave-vector $\bm k_\|=(k_x,k_y)$ is conserved for each scattering channel, it is also possible to use the following form
\begin{eqnarray}\label{S22}
{\rm \bm{H}}_{l,l} & = & \left[\begin{array}{cc}
\frac{k_{x}^{2}+k_{y}^{2}}{2m}+U_{l}-E_{F} & \Delta_{0}\cos(2\phi_{k})\Theta(l)\\
\Delta_{0}\cos(2\phi_{k})\Theta(l) & -(\frac{k_{x}^{2}+k_{y}^{2}}{2m}+U_{l}-E_{F})
\end{array}\right],
\end{eqnarray}
for the intra-layer part to make the calculation better compared with that of the continuum model. The two models (\ref{S20}) and (\ref{S22}) coincide in the long-wavelength limit (i.e. for small $k_x$ and $k_y$), with the correspondences $m=1/t$, $U_{l}=V_{l}-2t$, $\phi_{k}=\arg(k_{x}+ik_{y})$, and $\Delta_{0}=\Delta k_{\|}^{2}$.
We find that when $t^{\prime}<\varepsilon+E_F-U_{l}$, the Fermi surface of the S region (in its normal state) will take a cylinder-like shape.

\begin{figure}[H]
\centering\includegraphics[width=14cm]{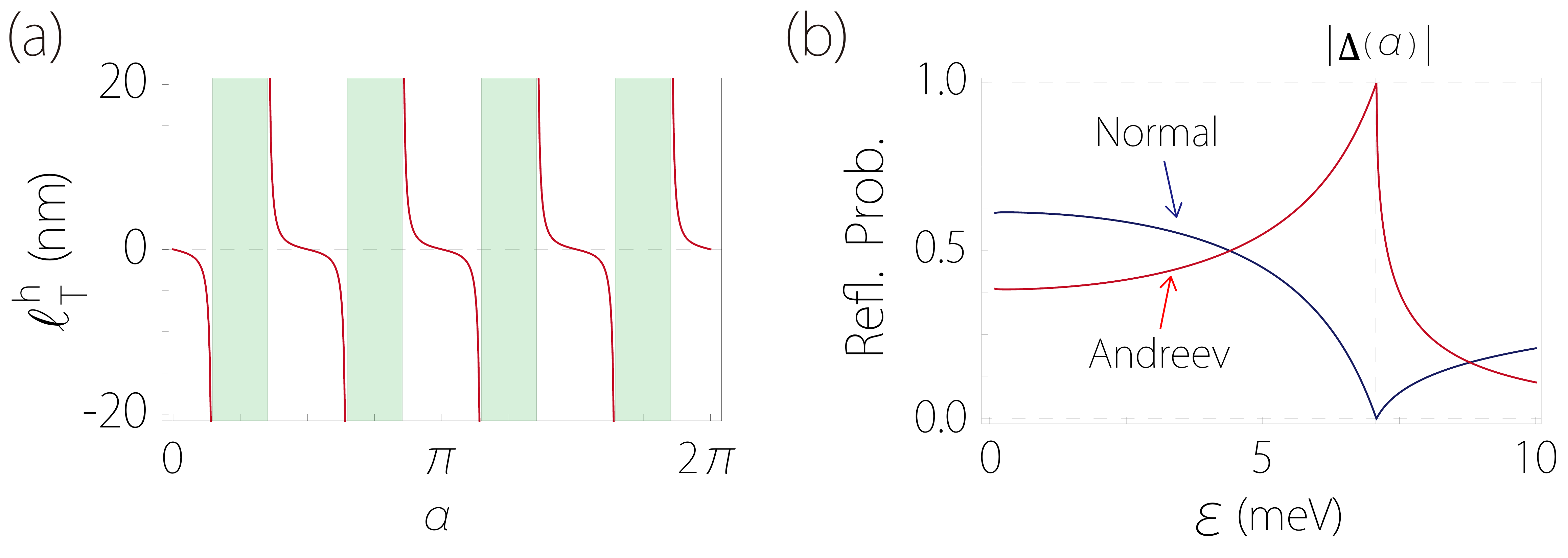}
\caption{
Results for $d_{x^2-y^2}$-wave pairing when the S side has a cylinder-like Fermi surface [see Fig.~4(e) of the main text].
(a) and (b) show the same quantities as in Fig.~4(b) and 4(d) of the main text.
In the calculation, we take the following parameters:
$t=1\ \rm{eV}$,  $t^{\prime}=0.07 t$, $t_{h}=(t+t^{\prime})/4$,  $U_{l}=t$ for $l\leq0$ and $U_{l}=0.35t+t^{\prime}$ for $l>0$,  $E_F=0.5\  \rm{eV}$,  and $\Delta_0=10\ \rm{meV}$.
We set $k_{\parallel}=0.2\  k_F^N$,  $\varepsilon=6\ \rm{meV}$ and the lattice constant as $4~{\rm {\AA}}$ in (a);  and $\alpha=\pi/8$ and $\gamma=\pi/10$ in (b).
In addition,
{ we have taken $t=1\ \rm{eV}$,  $t^{\prime}=0.04 t$, $E_F=0.5\  \rm{eV}$,  $U_{l}=t$ for $l\leq0$ and $U_{l}=0.3t+t^{\prime}$ for $l>0$,  $\Delta_0=20\ \rm{meV}$,  $\alpha=-\pi/6$,  $\varepsilon=9.8\ \rm{meV}$, and the lattice constant is set as $4~{\rm {\AA}}$ in Fig.~4(f) of the main text.}
\label{fig:S2}}
\end{figure}

Having constructed the block tridiagonal model, the scattering amplitudes between any incoming and outgoing propagating modes can be calculated using the iterative mode matching technique developed in Ref. \onlinecite{Ando} and Ref. \onlinecite{Khomyakov}. After obtaining the scattering amplitudes, the transverse shifts can be calculated using the formula presented in the main text. The results for the $d_{x^{2}-y^{2}}$-type pairing using the lattice model is shown in Fig.~\ref{fig:S2}.
Compared with the results for a closed ellipsoid Fermi surface [as in Fig.~4(a-d) of the main text], one observes that the key features [Features (i)-(iv) discussed in the main text] remain the same.

\section{Possible Experimental Detection}
\subsection{Transverse voltage signal at a structured NS junction}
Consider the setup shown in Fig.~4 in the main text. In this setup, the electrons injected from the N side have an \emph{average} velocity along the $x$-direction. The junction geometry is designed such that the flat interface (assumed to be the crystalline $a$-$b$ plane for the superconductor) makes a finite angle with respect to the $x$-direction. Hence the electron flow have an average oblique incident angle $\gamma$ when impinging on the interface. According to our discussion, on average, there will be a finite transverse shift along the $z$-direction in reflection. Assume that we are in the regime where the Andreev reflection dominates (which can be achieved, e.g., when $\varepsilon$ is close to the pairing gap). Then the reflected holes will have a finite average shift $\Delta \ell$ along the $z$-direction. This naturally leads to imbalanced charge accumulation between the top and bottom surfaces close to the interface on the N side, which can be detected as a voltage signal (see Fig.~4 in the main text).

Here, we provide a simple estimation for the induced voltage signal. Assume that the incident electrons have an average drift velocity $\bar{v}$ along $x$. In a time interval $\Delta t$, which may be taken on the order of the mean scattering time, the total number of electrons that impinge on the interface is given by
$
N=j S\Delta t,
$
where $j$ is the incident current density (we set the unit charge $e=1$) and  $S=w h$ is the cross-sectional area, with $w$ and $h$ being the width and height of the cross section. Assume that the scattering at the NS interface is dominated by Andreev reflection. The reflected holes have an average shift $\Delta\ell$ in the $z$-direction. This may be effectively modeled as an induced hole current in the $z$-direction:
$
I_{z}^h\approx N\Delta \ell/(h\Delta t)=j w\Delta \ell
$.
The current (and hence the surface charge accumulation) decays away from the interface due to scattering, so we may assume that this current is mainly distributed in a region within a mean free path $\lambda$ from the interface.

In this region, the carrier density distribution along the $z$-direction may be modeled by the diffusion equation (assuming that $h$ is much greater than the mean free path):
\begin{equation}
\frac{\partial n_\alpha}{\partial t}-D_\alpha\frac{\partial^2 n_\alpha}{\partial z^2}=-\frac{\partial j_z^\alpha}{\partial z}-\frac{n_\alpha}{\tau_\alpha},
\end{equation}
where $\alpha=e,h$ label the electrons and holes, $D_\alpha$ are the diffusion constants, and $\tau_\alpha$ are the quasiparticle lifetimes. For simplicity, we take the ansatz that $D_\alpha$ and $\tau_\alpha$ are independent of the carrier types. Let us first consider the top surface by assuming a semi-infinite sample extending for $z<0$. Consider a steady state and the current densities estimated from the above considerations are $j_z^e=0$, and $j_z^h(z)=j_z^h\Theta(-z)$ with
$
j_z^h\approx I_z^h /(\lambda w)= j\Delta \ell /\lambda .
$
By solving the diffusion equation, subjected to the boundary condition that $n(z=-\infty)=0$ for the net charge density $n=n_h-n_e$ due to charge neutrality in the bulk, we find that the net charge density near the surface is given by
\begin{equation}
n(z)=j_z^h \sqrt{\tau/D} e^{z/\sqrt{D\tau}}.
\end{equation}
The solution indicates that the charge accumulation is within a length scale of $\sqrt{D\tau}$ from the surface, with a total amount of charge $j_z^h \tau$ per unit area. Clearly, this charge accumulation is directly proportional to the shift $\Delta \ell$.

Similar analysis applies for the bottom surface, and the resulting charge accumulation has the opposite sign. This imbalanced charge accumulation between top and bottom surfaces can be detected as a voltage signal, which is on the order of
\begin{equation}
\Delta V\sim 2 n(z=0)/\mathcal{D},
\end{equation}
where $\mathcal{D}$ is the density of the states on the N side near the Fermi level. Admittedly, the treatment here is crude. An accurate determination of the charge accumulation profile would require more complicated simulations. Nevertheless, it offers a simplest estimation, and one expects that the key features of the result, such as $\Delta V\propto \Delta \ell$, would still hold.

For numerical estimations, we take the NS junction current density $j=10^{4}~ {\rm A\cdot cm^{-2}}$. With $\lambda=20$ nm, $\tau=10$ ps, $\Delta \ell=5$ nm,  $\sqrt{D\tau}= 40~{\rm nm}$ and { ${\cal D}=6 \times10^{19} {\rm cm^{-3}/eV}$}, typical for doped semiconductors such as GaAs \cite{Sotoodeh,Murzin,Ohno,Murakami}, one finds that the induced voltage can reach $\Delta V\sim 1$ mV. For nobel metals, $\Delta\ell$ would typically be smaller by an order of magnitude, and with $\lambda=50$ nm, $\tau=150$ ps,  $\sqrt{D\tau}= 70~{\rm nm}$ and { ${\cal D}=1.6\times10^{22} {\rm cm^{-3}/eV}$}, as appropriate for Ag \cite{Ashcroft,Gall,Bauer}, one finds that $\Delta V\sim 1\ \mu$V. With current technology, voltages can be measured with a resolution of 10 nV, so the induced voltage signal can be readily detected in experiment.

\subsection{Enhancing the shift in an SNS structure}

\begin{figure}[H]
\centering\includegraphics[width=11cm]{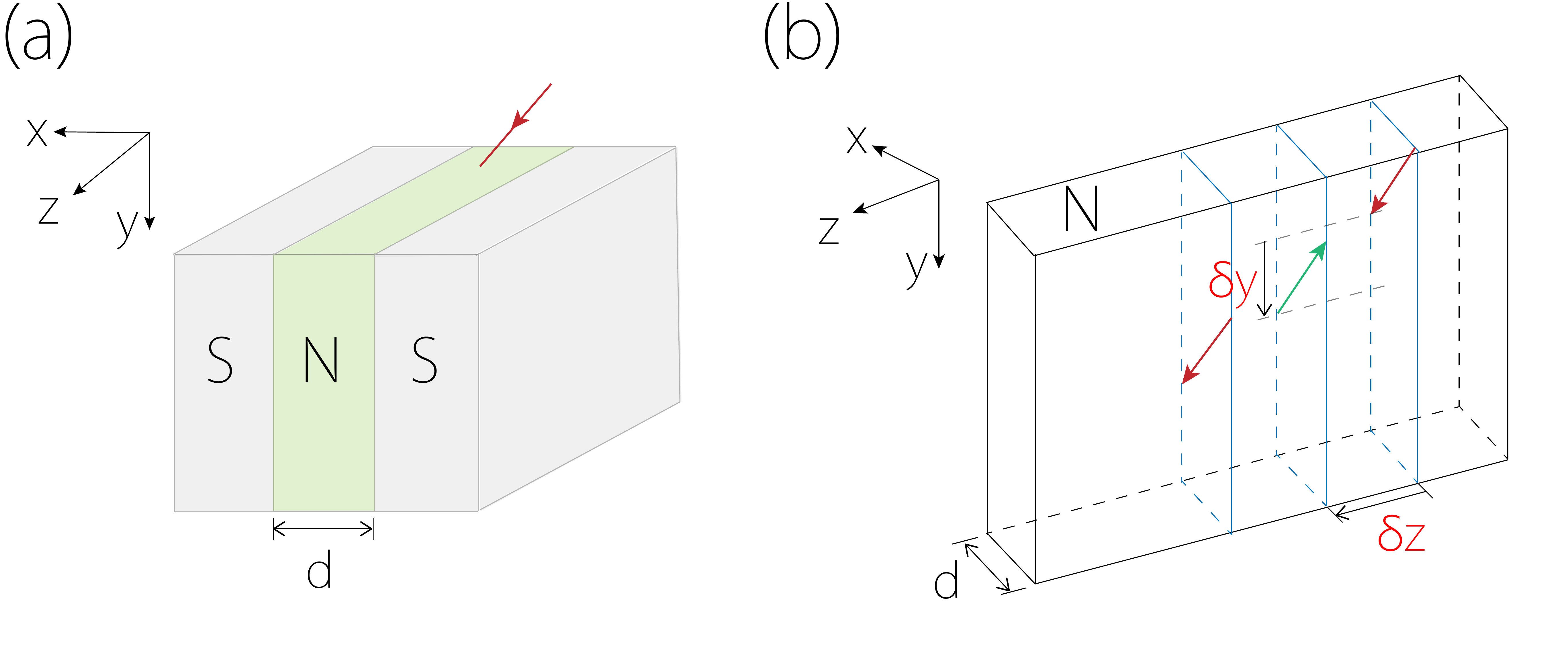}
\caption{
(a) Schematic of the setup with an SNS structure. A collimated electron beam injected into the N layer. Here we take the incident plane to be the $x$-$y$ plane. The beam undergoes multiple reflections between the two NS interfaces. (b) Illustration of the quasiparticle trajectory in the N region. In each reflection at the NS interface, the electron (or hole) will undergo a transverse shift ($\delta z$) and a longitudinal shift ($\delta y$).
\label{fig:Sdetect}}
\end{figure}

In this section, we suggest another setup to enhance the effect of transverse shift. It is an SNS structure as illustrated in in Fig.~\ref{fig:Sdetect}(a). A collimated electron beam injected into the structure (assume the incident plane is the $x$-$y$ plane) will undergo multiple reflections between the two NS interfaces [see Fig.~\ref{fig:Sdetect}(b)]. Here we assume that the interface scattering is dominated by Andreev reflection.
The shifts can then accumulate and lead to an anomalous velocity along the $z$-direction [see Fig.~\ref{fig:Sdetect}(b)]. Let $\delta z_i$ $(i=L,R)$ be the shift at the left and right interfaces. Then the anomalous velocity along $z$ is given by
\begin{equation}
v^a_z=\frac{\delta z_L+\delta z_R}{2\delta t},
\end{equation}
where
\begin{equation}
\delta t=\frac{d}{v\cos\gamma}
\end{equation}
is the time for the travel between the two interfaces. Here, $d$ is the thickness of the N region, $\gamma$ is the incident angle, and $v\sim v_F$ is the particle velocity.

If we assume that the two S regions are with the same chiral $p$-wave pairings, and we take $d=20$ nm and $\gamma=\pi/4$, for both Ag and doped GaAs (taking $E_F=5.5$ eV, $m=m_e$ for Ag \cite{Ashcroft}; $E_F=0.26$ eV, $m=0.065m_e$ for doped GaAs \cite{Sotoodeh}), we find that $v^a_z\sim 10^4$ m/s, which is quite sizable. This anomalous velocity will lead to enhanced charge accumulation on the two opposite surfaces perpendicular to the $z$-direction, which can be detected as voltage signals.

We mention that there is also an anomalous velocity along $y$, which is due to the longitudinal shift in the scattering plane [as illustrated in Fig.~\ref{fig:Sdetect}(b)]. The detailed analysis of this longitudinal shift will be presented elsewhere.
